\documentclass{emulateapj}
\usepackage{lscape}

\shorttitle{Abundances in the Sculptor dSph}
\shortauthors{Kirby et al.}
\slugcomment{Accepted to ApJ on 2009 Sep 15}

\begin{document}
\newcommand{\teff}{$T_{\rm{eff}}$}
\newcommand{\mathteff}{T_{\rm eff}}
\newcommand{\logg}{$\log g$}
\newcommand{\mathlogg}{\log g}
\newcommand{\feh}{[Fe/H]}
\newcommand{\mathfeh}{{\rm [Fe/H]}}
\newcommand{\afe}{[$\alpha$/Fe]}
\newcommand{\mathafe}{{\rm [\alpha/Fe]}}
\newcommand{\ah}{[$\alpha$/H]}
\newcommand{\mathah}{{\rm [\alpha/H]}}
\newcommand{\vt}{$v_t$}
\newcommand{\mathvt}{v_t}

\newcommand{\sclnbad}{2}
\newcommand{\sclngood}{410}
\newcommand{\sclndup}{17}
\newcommand{\sclnunique}{393}
\newcommand{\fehsyserr}{0.136}
\newcommand{\alphasyserr}{0.102}
\newcommand{\mgfesyserr}{0.108}
\newcommand{\sifesyserr}{0.179}
\newcommand{\cafesyserr}{0.087}
\newcommand{\tifesyserr}{0.101}
\newcommand{\scllowv}{84.8}
\newcommand{\sclhighv}{138.3}
\newcommand{\sclmeanv}{111.6}
\newcommand{\sclmeanverr}{0.5}
\newcommand{\sclvdupsigma}{3.9}
\newcommand{\sclsigmav}{8.0}
\newcommand{\sclsigmaverr}{0.7}
\newcommand{\sclmeanvother}{110.4}
\newcommand{\sclmeanverrother}{0.8}
\newcommand{\sclsigmavother}{8.8}
\newcommand{\sclsigmaverrother}{0.6}
\newcommand{\sclmeanvdiff}{+1.2}
\newcommand{\sclsigmavdiff}{-0.9}
\newcommand{\sclmeanvw}{111.3}
\newcommand{\sclmeanverrw}{0.2}
\newcommand{\sclsigmavw}{8.7}
\newcommand{\sclsigmaverrw}{0.2}
\newcommand{\sclmeanvdiffw}{+0.6}
\newcommand{\sclsigmavdiffw}{-1.0}
\newcommand{\sclalphaerfsigma}{1.5}
\newcommand{\sclnnonmember}{5}
\newcommand{\sclnmember}{388}
\newcommand{\sclnvmp}{80}
\newcommand{\sclmedtefferrrand}{98}
\newcommand{\sclmedtefferrsys}{58}
\newcommand{\sclmedtefferrtot}{117}
\newcommand{\sclmedloggerrrand}{0.06}
\newcommand{\sclmedloggerrsys}{0.01}
\newcommand{\sclmedloggerrtot}{0.06}
\newcommand{\sclfehhrssigma}{0.14}
\newcommand{\sclfehbatsigma}{0.14}
\newcommand{\sclcafebatsigma}{0.20}
\newcommand{\sclcafebatcorr}{0.53}
\newcommand{\sclnbat}{47}
\newcommand{\sclmdfhelmean}{-1.82}
\newcommand{\sclmdfhelsigma}{0.35}
\newcommand{\sclmdfbatmean}{-1.56}
\newcommand{\sclmdfbatsigma}{0.38}
\newcommand{\sclnbathrs}{7}
\newcommand{\sclfehbathrssigma}{0.16}
\newcommand{\sclfehcuthrssigma}{0.14}
\newcommand{\sclfehslope}{-0.030}
\newcommand{\sclfehslopeerr}{0.003}
\newcommand{\sclfehslopekpc}{-1.21}
\newcommand{\sclfehslopekpcerr}{0.13}
\newcommand{\sclafeslope}{+0.013}
\newcommand{\sclafeslopeerr}{0.003}
\newcommand{\sclafeslopekpc}{+0.54}
\newcommand{\sclafeslopekpcerr}{0.11}
\newcommand{\sclalphafehrssigma}{0.13}
\newcommand{\sclfehrange}{NaN}
\newcommand{\sclfehinitial}{-2.92}
\newcommand{\sclfehinitialerr}{0.00}
\newcommand{\sclfehfinal}{     }
\newcommand{\sclfehfinalerr}{0.00}
\newcommand{\sclsimpleyield}{0.031}
\newcommand{\sclsimpleyielderr}{0.001}
\newcommand{\sclinfallm}{1.76}
\newcommand{\sclinfallyield}{0.035}
\newcommand{\sclprobratio}{2.74}
\newcommand{\sclfehspread}{0.41}
\newcommand{\sclmgfespread}{0.10}
\newcommand{\sclsifespread}{0.11}
\newcommand{\sclcafespread}{0.10}
\newcommand{\scltifespread}{0.14}
\newcommand{\sclalphafespread}{0.16}
\newcommand{\sclmgfespreadsub}{-0.13}
\newcommand{\sclsifespreadsub}{-0.15}
\newcommand{\sclcafespreadsub}{-0.06}
\newcommand{\scltifespreadsub}{0.08}
\newcommand{\sclalphafespreadsub}{0.04}
\newcommand{\sclfehmean}{-1.58}
\newcommand{\sclfehsigma}{0.41}
\newcommand{\sclfehmedian}{-1.58}
\newcommand{\sclfehmad}{0.33}
\newcommand{\sclfehiqr}{0.67}
\newcommand{\sclempfehone}{-3.01}
\newcommand{\sclempfeherrone}{0.32}
\newcommand{\sclempvone}{17.94}
\newcommand{\sclempfehtwo}{-3.80}
\newcommand{\sclempfeherrtwo}{0.28}
\newcommand{\sclempvtwo}{18.19}
\newcommand{\sclcmdfeh}{-1.31}

\title{Multi-Element Abundance Measurements from Medium-Resolution
  Spectra. \\ I. The Sculptor Dwarf Spheroidal Galaxy}

\author{Evan~N.~Kirby, Puragra~Guhathakurta, Michael Bolte}
\affil{University of California Observatories/Lick
  Observatory, Department of Astronomy \& Astrophysics, \\ 
  University of California, Santa Cruz, CA 95064}

\author{Christopher~Sneden}
\affil{McDonald Observatory, University of Texas, Austin, TX 78712}

\and

\author{Marla~C.~Geha}
\affil{Astronomy Department, Yale University, New Haven, CT 06520}

\altaffiltext{1}{Data herein were obtained at the W.~M. Keck
  Observatory, which is operated as a scientific partnership among the
  California Institute of Technology, the University of California,
  and NASA.  The Observatory was made possible by the generous
  financial support of the W.~M. Keck Foundation.}



\begin{abstract}

We present measurements of Fe, Mg, Si, Ca, and Ti abundances for
\sclnmember\ radial velocity member stars in the Sculptor dwarf
spheroidal galaxy (dSph), a satellite of the Milky Way.  This is the
largest sample of individual $\alpha$ element (Mg, Si, Ca, Ti)
abundance measurements in any single dSph.  The measurements are made
from Keck/DEIMOS medium-resolution spectra (6400--9000~\AA, $R \sim
6500$).  Based on comparisons to published high-resolution ($R \ga
20000$) spectroscopic measurements, our measurements have
uncertainties of $\sigma\mathfeh = \sclfehhrssigma$ and
$\sigma\mathafe = \sclalphafehrssigma$.  The Sculptor \feh\
distribution has a mean $\langle\mathfeh\rangle = \sclfehmean$ and is
asymmetric with a long, metal-poor tail, indicative of a history of
extended star formation.  Sculptor has a larger fraction of stars with
$\mathfeh < -2$ than the Milky Way halo.  We have discovered one star
with $\mathfeh = \sclempfehtwo \pm \sclempfeherrtwo$, which is the
most metal-poor star known anywhere except the Milky Way halo, but
high-resolution spectroscopy is needed to measure this star's detailed
abundances.  As has been previously reported based on high-resolution
spectroscopy, \afe\ in Sculptor falls as \feh\ increases.  The
metal-rich stars ($\mathfeh \sim -1.5$) have lower \afe\ than Galactic
halo field stars of comparable metallicity.  This indicates that star
formation proceeded more gradually in Sculptor than in the Galactic
halo.  We also observe radial abundance gradients of $\sclfehslope \pm
\sclfehslopeerr$~dex per arcmin in \feh\ and $\sclafeslope \pm
\sclafeslopeerr$~dex per arcmin in \afe\ out to 11 arcmin (275~pc).
Together, these measurements cast Sculptor and possibly other
surviving dSphs as representative of the dwarf galaxies from which the
metal-poor tail of the Galactic halo formed.

\end{abstract}


\section{Introduction}
\label{sec:intro}

The dwarf spheroidal galaxy (dSph) companions of the Milky Way (MW)
are excellent laboratories for investigating the chemical evolution
and star formation histories of dwarf galaxies.  These galaxies have
undergone at most a few star formation episodes \citep{hol06} and are
dynamically simple \citep{wal07}.  The dSphs of the MW provide an
opportunity to examine closely the processes that establish the galaxy
luminosity-metallicity relation \citep[e.g.,][]{sal09}.


The MW dSphs are also considered to be strong candidates of a
population of dwarf galaxies that were tidally stripped by the young
Galaxy and eventually incorporated into the Galactic halo.  This
scenario has become central to our picture of how large galaxies form
\citep{sea78,rob05}.  Important tests of this scenario are to compare
the details of the metallicity distribution function of the collection
of dSphs to that of the Galactic halo stars and to compare abundance
ratio patterns seen in dSphs to those measured for the halo
\citep[e.g.,][]{ven04}.

To date, each of these areas has been hampered by the small sample of
dSph stars for which high-quality measurements of [Fe/H] and abundance
ratios for other elements have been available.  \citet{lan04} compared
their models of dSphs less massive than Sagittarius to six or fewer
stars per galaxy.  The usual approach for high-quality detailed
abundance determinations is to use high-resolution spectroscopy (HRS,
$R > 20000$) of individual stars.  Because of the large distances to
even the nearest dSphs, these are time-consuming observations even
using the largest telescopes.

Our approach is to derive abundances from medium-resolution
spectroscopy (MRS, $R \sim 6500$) using the Deep Imaging Multi-Object
Spectrometer \citep[DEIMOS,][]{fab03} on the Keck~II telescope.  As
demonstrated by \citet{kir08a,kir08b}, accurate measurements can be
made for Fe and some $\alpha$ elements (Mg, Si, Ca, and Ti) with these
individual stellar spectra.  \citet{she09} demonstrated similarly
precise results using the Keck~I LRIS spectrometer on a sample of
individual stars in the \object[NAME Leo II dSph]{Leo~II dSph}.  In a
typical dSph, the DEIMOS field of view allows between 80 and 150 red
giant stars to be targeted per multi-object mask.  Samples of several
hundred giants can be observed in a given dSph.  The Dwarf Abundances
and Radial Velocities team \citep[DART,][hereafter T04]{tol04} has
been collecting a combination of MRS and HRS in dSphs to exploit the
advantages of both techniques.

This paper is the first in a series that explores the multi-element
abundances of stellar systems measured with MRS.  The particular focus
of this series is to characterize the distributions of \feh\ and \afe\
in MW dSphs.  These measurements will provide insight into the role of
dSphs in building the Galactic stellar halo
\citep[i.e.,][]{sea78,whi78}.

Our first target is the \object[NAME SCULPTOR dSph]{Sculptor dSph}
  \citep[$\alpha = 1^{\rm{h}}00^{\rm{m}}$, $\delta = -33^{\circ}43'$,
  $M_V = -11.1$,][]{mat98}.  Sculptor has been a favored HRS and MRS
  target for the past ten years.  Of all the dSphs, it appears most
  often in explanations of dSph chemical evolution and galaxy
  formation \citep[e.g., T04,][]{she03,gei07}.  T04 discovered that
  Sculptor is actually ``two galaxies'' in one, with two stellar
  populations that are kinematically and compositionally distinct.
  \citet{bat06} later showed that \object[NAME FORNAX dSph]{Fornax}
  also displays multiple stellar populations with different
  kinematics, spatial extents, and metallicities.  But Sculptor is
  also unique in that it is the only MW dSph known to rotate
  \citep{bat08a}.  Recently, \citet{wal09} published radial velocities
  for 1365 Sculptor members, and \citet{ven05,ven08} presented
  high-resolution abundance measurements of Mg, Ca, Ti, and Fe for 91
  stars in Sculptor.  They also measured Y, Ba, and Eu for some of
  those stars.

This paper consists of six sections and an appendix.
Section~\ref{sec:obs} introduces the spectroscopic target selection
and observations, and Sec.~\ref{sec:prep} explains how the spectra are
prepared for abundance measurements.  Section~\ref{sec:measure}
describes the technique to extract abundances, which builds on the
method described by \citet*[][hereafter KGS08]{kir08a}.  In
Sec.~\ref{sec:abund}, we present the metallicity distribution and
multi-element abundance trends of Sculptor.  In Sec.~\ref{sec:concl},
we summarize our findings in the context of dSph chemical evolution
and the formation of the Galaxy.  Finally, we devote the appendix to
quantifying the uncertainties in our MRS measurements, including
comparisons to independent HRS of the same stars.


\section{Observations}
\label{sec:obs}

\subsection{Target Selection}

We selected targets from the Sculptor photometric catalog of
\citet{wes06}.  The catalog includes photometry in three filters: $M$
and $T_2$ in the Washington system, and the intermediate-width DDO51
filter (henceforth called $D$) centered at 5150~\AA.  This band probes
the flux from a spectral region susceptible to absorption by the
surface gravity-sensitive \ion{Mg}{1} and MgH lines.  \citet{maj00}
and \citet{wes06} outlined the procedure for distinguishing between
distant red giant stars and foreground Galactic dwarf stars using
these three filters.  We followed the same procedure to select a
sample of red giant candidates from the Sculptor $MT_2D$ catalog.

\begin{deluxetable*}{lcccccccc}
\tablewidth{0pt}
\tablecolumns{6}
\tablecaption{Targets with Previous High-Resolution Abundances\label{tab:hrslist}}
\tablehead{\colhead{name} & \colhead{reference} & \colhead{RA} & \colhead{Dec} & \colhead{$M$} & \colhead{$T_2$}}
\startdata
H482 & \citet{she03} & $00^{\mathrm{h}} 59^{\mathrm{m}} 58 \fs 2$ & $-33 \arcdeg 41 \arcmin 08 \arcsec$ & $17.967 \pm 0.030$ & $16.324 \pm 0.020$ \\
H459 & \citet{she03} & $01^{\mathrm{h}} 00^{\mathrm{m}} 12 \fs 5$ & $-33 \arcdeg 43 \arcmin 01 \arcsec$ & $18.465 \pm 0.032$ & $16.924 \pm 0.031$ \\
H479 & \citet{she03} & $01^{\mathrm{h}} 00^{\mathrm{m}} 12 \fs 7$ & $-33 \arcdeg 41 \arcmin 15 \arcsec$ & $17.562 \pm 0.023$ & $15.860 \pm 0.030$ \\
H400 & \citet{she03} & $01^{\mathrm{h}} 00^{\mathrm{m}} 17 \fs 0$ & $-33 \arcdeg 45 \arcmin 13 \arcsec$ & $18.413 \pm 0.030$ & $17.140 \pm 0.027$ \\
H461 & \citet{she03} & $01^{\mathrm{h}} 00^{\mathrm{m}} 18 \fs 2$ & $-33 \arcdeg 42 \arcmin 12 \arcsec$ & $17.806 \pm 0.028$ & $16.166 \pm 0.027$ \\
1446 & \citet{gei05} & $00^{\mathrm{h}} 59^{\mathrm{m}} 46 \fs 4$ & $-33 \arcdeg 41 \arcmin 23 \arcsec$ & $17.618 \pm 0.023$ & $15.695 \pm 0.022$ \\
195  & \citet{gei05} & $00^{\mathrm{h}} 59^{\mathrm{m}} 55 \fs 6$ & $-33 \arcdeg 46 \arcmin 39 \arcsec$ & $17.515 \pm 0.022$ & $15.845 \pm 0.018$ \\
982  & \citet{gei05} & $01^{\mathrm{h}} 00^{\mathrm{m}} 16 \fs 2$ & $-33 \arcdeg 42 \arcmin 37 \arcsec$ & $17.433 \pm 0.025$ & $15.552 \pm 0.028$ \\
770  & \citet{gei05} & $01^{\mathrm{h}} 00^{\mathrm{m}} 23 \fs 8$ & $-33 \arcdeg 42 \arcmin 17 \arcsec$ & $17.623 \pm 0.025$ & $15.857 \pm 0.026$ \\
\enddata
\end{deluxetable*}

Nine stars, listed in Table~\ref{tab:hrslist}, have previously
published HRS abundance measurements \citep{she03,gei05}.  These stars
were observed and provide the basis for demonstrating the accuracy of
the MRS abundance measurements, described in the appendix.

\subsection{Slitmask Design}

We designed the DEIMOS slitmasks with the IRAF software module
\texttt{dsimulator}.\footnote{\url{http://www.ucolick.org/$^\sim$phillips/deimos\_ref/masks.html}}
Each slitmask subtended approximately $16' \times 4'$.  In order to
adequately subtract night sky emission lines, we required a minimum
slit length of $4''$.  The minimum space between slits was $0 \farcs
35$.  When these constraints forced the selection of one among
multiple possible red giant candidates, the brightest object was
selected.  The slits were designed to be at the approximate
parallactic angle at the anticipated time of observation
($-25^{\circ}$).  This choice minimized the small light losses due to
differential atmospheric refraction.  This configuration was
especially important for Sculptor, which was visible from Keck
Observatory only at a low elevation.  The slitmasks' sky position
angle (PA) was $-35^{\circ}$.  The $10^{\circ}$ offset between the
slit PA and the slitmask PA tilted the night sky emission lines
relative to the CCD pixel grid to increase the subpixel wavelength
sampling and improve sky subtraction.

\begin{figure}[t!]
\includegraphics[width=\linewidth]{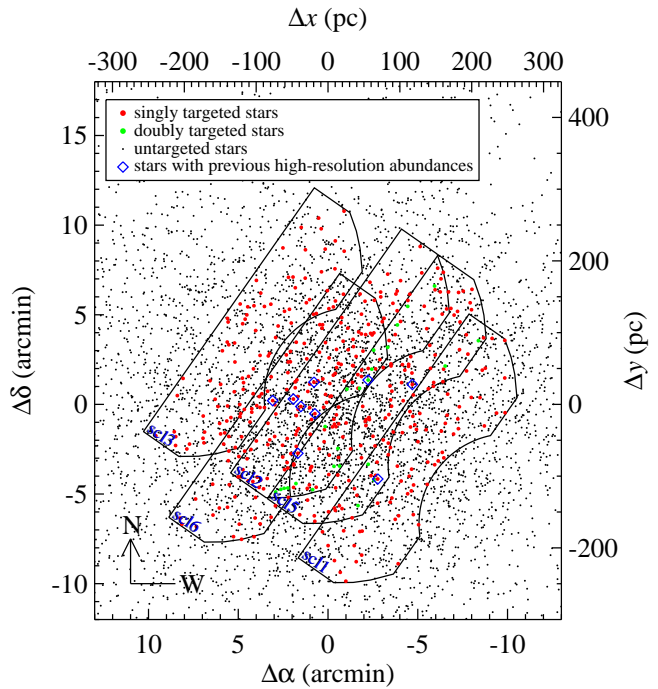}
\caption{DEIMOS slitmask footprints laid over a map of sources from
  the photometric catalog.  Targets selected for spectroscopy are
  shown in red.  Targets observed in more than one mask are shown in
  green.  Blue diamonds enclose stars with previous HRS abundance
  measurements.  The left and bottom axis scales show the angular
  displacement in arcmin from the center of the galaxy
  \citep[$\alpha_0 = 1^{\rm{h}}00^{\rm{m}}09^{\rm{s}}$, $\delta_0 =
    -33^{\circ}42'30''$,][]{mat98}, and the right and top axis scales
  show the physical displacement for an assumed distance of 85.9~kpc
  \citep{pie08}.\label{fig:coords}}
\end{figure}

Figure~\ref{fig:coords} shows the coordinates of all the objects in
the catalog regardless of their probability of membership in Sculptor.
Five DEIMOS slitmask footprints enclose the spectroscopic targets:
scl1, scl2, scl3, scl5, and scl6 (see Tab.~\ref{tab:obs}).  The scl5
slitmask included 24 targets also included on other masks.  These
duplicate observations provide estimates of uncertainty in radial
velocity and abundance measurements (Sec.~\ref{sec:velocities} and
Sec.~\ref{sec:duplicate}).  The spectral coverage of each slit is not
the same.  The minimum and maximum wavelengths of spectra of targets
near the long, straight edge of the DEIMOS footprint can be up to
400~\AA\ lower than for targets near the irregularly shaped edge of
the footprint (upper left and lower right of the slitmask footprints
in Fig.~\ref{fig:coords}, respectively).  Furthermore, spectra of
targets near either extreme of the long axis of the slitmask suffered
from vignetting which reduced the spectral range.  It is important to
keep these differences of spectral range in mind when interpreting the
differences of measurements derived from duplicate observations.

\begin{figure}[t!]
\includegraphics[width=\linewidth]{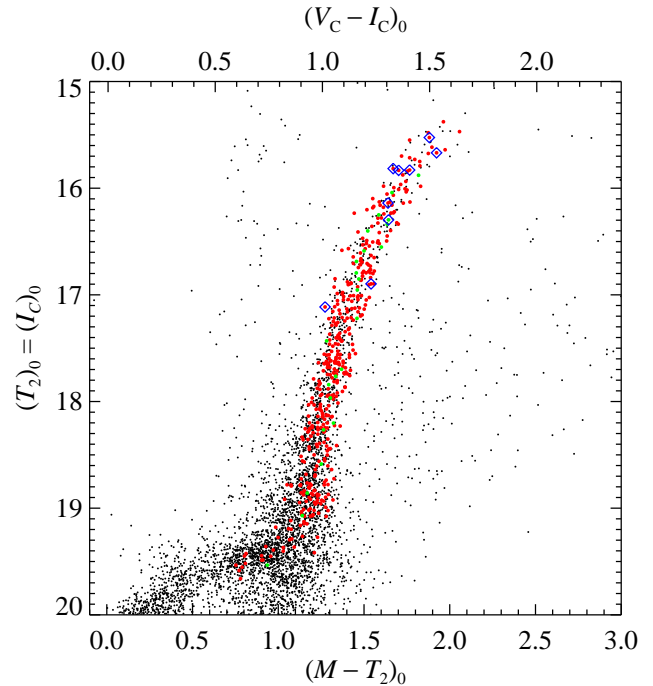}
\caption{Color-magnitude diagram in the Washington and Cousins systems
  for the sources within the right ascension and declination ranges
  shown in Fig.~\ref{fig:coords}.  The symbols have the same meanings
  as in Fig.~\ref{fig:coords}.  The transformation from the Washington
  system ($M$ and $T_2$) to the Cousins system ($V_{\rm C}$ and
  $I_{\rm C}$) is $I_{\rm C} = T_2$ and $V_{\rm C} - I_{\rm C} =
  0.800(M-T_2) - 0.006$ \citep{maj00}.\label{fig:cmd}}
\end{figure}

Figure~\ref{fig:cmd} shows the color-magnitude diagram (CMD) of the
targets within the right ascension and declination ranges of the axes
in Fig.~\ref{fig:coords}.  The $MT_2D$ membership criteria caused the
selected red giants to form a tight sequence.  This selection may have
imposed a metallicity bias on the spectroscopic sample.  Although only
a tiny fraction of stars lay outside the main locus of the red giant
branch, some may have been spectroscopically untargeted members of
Sculptor.  For example, if Sculptor contained any old stars with
$\mathfeh \ga -0.5$, they would have been too red to be included in
the spectroscopic sample.  Any such metallicity bias should have
excluded at most a few stars.

\subsection{Spectroscopic Configuration and Exposures}

\begin{deluxetable}{lcccc}
\tablewidth{0pt}
\tablecolumns{5}
\tablecaption{DEIMOS Observations\label{tab:obs}}
\tablehead{\colhead{Slitmask} & \colhead{Targets} & \colhead{UT Date} & \colhead{Exposures} & \colhead{Seeing}}
\startdata
scl1    & \phn86  & 2008 Aug 3\phn & $3 \times 1200$~s    & $0 \farcs 8$ \\
scl2    &    106  & 2008 Aug 3\phn & $2 \times 900$~s\phn & $0 \farcs 8$ \\
scl3    & \phn87  & 2008 Aug 4\phn & $1 \times 462$~s\phn & $0 \farcs 9$ \\
        &         & 2008 Aug 31    & $1 \times 1000$~s    & $0 \farcs 8$ \\
        &         & 2008 Aug 31    & $1 \times 834$~s\phn & $0 \farcs 8$ \\
scl5    & \phn95  & 2008 Sep 1\phn & $3 \times 720$~s\phn & $0 \farcs 8$ \\
scl6    & \phn91  & 2008 Sep 1\phn & $3 \times 720$~s\phn & $1 \farcs 2$ \\
\enddata
\tablecomments{The scl4 slitmask was not observed.}
\end{deluxetable}

Our observing strategy was nearly identical to that of \citet{sim07}
and \citet{kir08a}.  In summary, we used with the 1200 lines~mm$^{-1}$
grating at a central wavelength of 7800~\AA.  The slit widths were
$0\farcs 7$, yielding a spectral resolution of $\sim 1.3$~\AA\ FWHM
(resolving power $R \sim 6500$ at 8500~\AA).  The OG550 filter blocked
diffraction orders higher than $m=1$.  The spectral range was about
6400--9000~\AA\ with variation depending on the slit's location along
the dispersion axis.  Exposures of Kr, Ne, Ar, and Xe arc lamps
provided wavelength calibration, and exposures of a quartz lamp
provided flat fielding.  Table~\ref{tab:obs} lists the number of
targets for each slitmask, the dates of observations, the exposure
times, and the approximate seeing.


\section{Data Reduction}
\label{sec:prep}

\subsection{Extraction of One-Dimensional Spectra}

We reduced the raw frames using version 1.1.4 of the DEIMOS data
reduction pipeline developed by the DEEP Galaxy Redshift
Survey.\footnote{\url{http://astro.berkeley.edu/$^{\sim}$cooper/deep/spec2d/}}
\citet{guh06} give the details of the data reduction.  We also made
use of the optimizations to the code described by \citet[Sec.~2.2 of
  their article]{sim07}.  These modifications provided better
extraction of unresolved stellar sources.

In summary, the pipeline traced the edges of slits in the flat field
to determine the CCD location of each slit.  The wavelength solution
was given by a polynomial fit to the CCD pixel locations of arc lamp
lines.  Each exposure of stellar targets was rectified and then
sky-subtracted based on a B-spline model of the night sky emission
lines.  Next, the exposures were combined with cosmic ray rejection
into one two-dimensional spectrum for each slit.  Finally, the
one-dimensional stellar spectrum was extracted from a small spatial
window encompassing the light of the star in the two-dimensional
spectrum.  The product of the pipeline was a wavelength-calibrated,
sky-subtracted, cosmic ray-cleaned, one-dimensional spectrum for each
target.

Some of the spectra suffered from unrecoverable defects, such as a
failure to find an acceptable polynomial fit to the wavelength
solution.  There were 53 such spectra.  An additional \sclnbad\ spectra
had such poor signal-to-noise ratios (SNR) that abundance measurements
were impossible, leaving \sclngood\ useful spectra, comprising \sclnunique\
unique targets and \sclndup\ duplicate measurements.

\begin{figure}[t!]
\includegraphics[width=\linewidth]{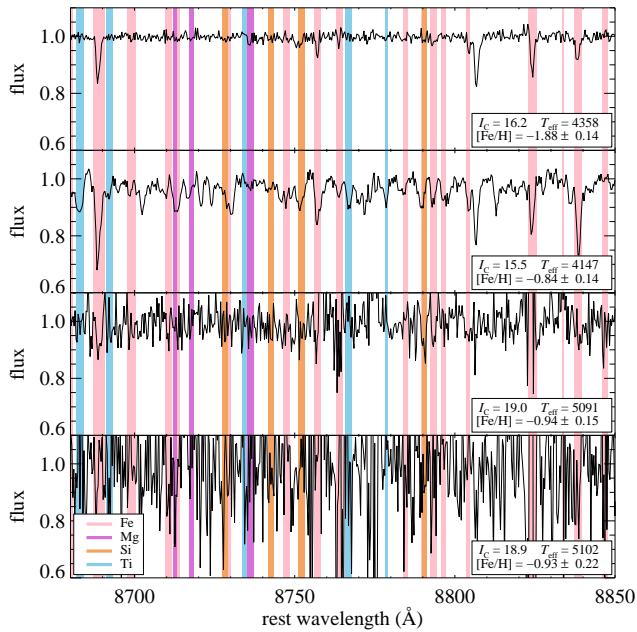}
\caption{Examples of small regions of DEIMOS spectra of four different
  stars.  The continuum in each spectrum has been normalized to unity.
  The $I_{\rm C}$ magnitude, measured effective temperature, and
  measured \feh\ is given for each star.  The top two panels show two
  stars with very different \feh, and the bottom two panels show two
  stars with nearly the same temperature and \feh\ but different SNR.
  The colors show the regions used to measure each of the Fe, Mg, Si,
  Ca, and Ti abundances (see
  Fig.~\ref{fig:coadd}).\label{fig:snexamples}}
\end{figure}

Figure~\ref{fig:snexamples} shows four example spectra at a variety of
$I_{\rm C}$ magnitudes, effective temperatures, and \feh.  The two
upper panels show stars in the top 10\% of the SNR distribution.  The
two lower panels show stars from the middle and bottom 10\% of the
distribution.

The one-dimensional DEIMOS spectra needed to be prepared for abundance
measurements.  The preparation included velocity measurement, removal
of telluric absorption, and continuum division.  KGS08 (their Sec.~3)
described these preparations in detail.  We followed the same process
with some notable exceptions, described below.

\subsection{Telluric Absorption Correction}

We removed the absorption introduced into the stellar spectra by the
earth's atmosphere in the same manner as KGS08: division by a hot star
template spectrum.  However, the high airmass of the Sculptor
observations caused much stronger absorption than KGS08 observed in
globular cluster (GC) spectra.  Even after scaling the hot star
template spectrum by the airmass, large residuals in the Sculptor
stellar spectra remained.  Consequently, we masked spectral regions of
heavy telluric absorption before measuring abundances.  These regions
are 6864--6932~\AA, 7162--7320~\AA, 7591--7703~\AA, 8128--8351~\AA,
and 8938--10000~\AA\ (see Fig.~\ref{fig:coadd}).

\subsection{Radial Velocities and Spectroscopic Membership Determination}
\label{sec:velocities}

Our primary interest in this paper is chemical abundances, and we
measured radial velocities only to determine membership and to shift
the spectra into the rest frame.

Following KGS08, we measured stellar radial velocities by
cross-correlation with a template spectrum.  However, KGS08
cross-correlated the observed spectra against synthetic spectra
whereas we cross-correlated the observed spectra against high SNR
template spectra of stars observed with DEIMOS.  Templates observed
with the same instrument should provide more accurate radial velocity
measurements than synthetic templates.  \citet{sim07} provided their
template spectra to us.  For the rest of the analysis, the spectra are
shifted to the rest frame.

\begin{figure}[t!]
\includegraphics[width=\linewidth]{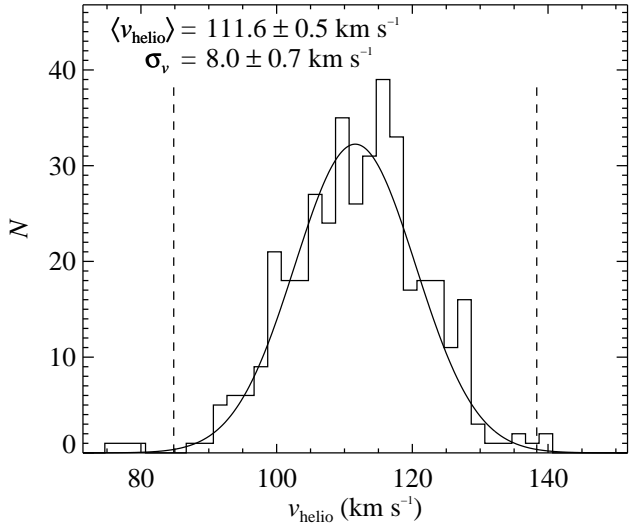}
\caption{Distribution of measured radial velocities for targets in the
  Sculptor field along with the best-fit Gaussian.  The top left label
  gives the mean and standard deviation of this Gaussian fit.  The
  five stars outside of the dashed lines are not considered Sculptor
  members.  The velocity range of this plot includes all stars for
  which a velocity measurement was possible.\label{fig:vhist}}
\end{figure}

Although the $MT_2D$ selection eliminated almost all of the foreground
MW contaminants from the spectroscopic sample, we checked the
membership of each target by radial velocity selection.
Figure~\ref{fig:vhist} shows the distribution of radial velocities in
this spectroscopic data set along with the best-fit Gaussian.  We
consider the radial velocity limits of Sculptor membership to be
$\scllowv$~km~s$^{-1} < v_r < \sclhighv$~km~s$^{-1}$.  We chose these limits
because beyond them, the expected number of Sculptor members per
2~km~s$^{-1}$ bin \citep[the approximate maximum velocity resolution
of DEIMOS,][]{sim07} is fewer than 0.5.  This selection eliminated
just \sclnnonmember\ out of \sclnunique\ unique targets.

As a check on our procedure, we compared some derived quantities from
the velocity distribution to previous measurements.  The mean velocity
of our sample is $\langle v_{\rm{helio}}\rangle = \sclmeanv \pm
\sclmeanverr$~km~s$^{-1}$ with a dispersion of $\sigma_v = \sclsigmav
\pm \sclsigmaverr$~km~s$^{-1}$.  The velocity dispersion is the
per-measurement velocity error subtracted in quadrature from the
$1\sigma$ width of the velocity distribution.  The per-measurement
error is \sclvdupsigma~km~s$^{-1}$, which is the standard deviations
of the differences in measured velocities for the \sclndup\ duplicate
spectra.  In comparison, \citet{wes06} found $\langle
v_{\rm{helio}}\rangle = \sclmeanvother \pm
\sclmeanverrother$~km~s$^{-1}$ (difference of $\sclmeanvdiff\sigma$)
and $\sigma_v = \sclsigmavother \pm \sclsigmaverrother$~km~s$^{-1}$
(difference of $\sclsigmavdiff\sigma$).  The comparison of the
velocity dispersions depends on the assumed binary fraction
\citep{que95} and---given the presence of multiple kinematically and
spatially distinct populations in Sculptor (T04)---the region of
spectroscopic selection.  Furthermore, \citet{wal07,wal09} reported
velocity dispersion gradients, and \citet{bat08a} reported mean
velocity gradients along the major axis, indicating rotation.  We
choose not to address the kinematic complexity of this system in this
paper.

\subsection{Continuum Determination}

In the abundance analysis described in Sec.~\ref{sec:measure}, it is
necessary to normalize each stellar spectrum by dividing by the slowly
varying stellar continuum.  KGS08 determined the continuum by
smoothing the regions of the stellar spectrum free from strong
absorption lines.  Instead of smoothing, we fit a B-spline with a
breakpoint spacing of 150~\AA\ to the same ``continuum regions''
defined by KGS08.  Each pixel was weighted by its inverse variance in
the fit.  Furthermore, the fit was performed iteratively such that
pixels that deviated from the fit by more than $5\sigma$ were removed
from the next iteration of the fit.

The spline fit results in a smoother continuum determination than
smoothing.  Whereas the smoothed continuum value may be influenced
heavily by one or a few pixels within a relatively small smoothing
kernel, the spline fit is a global fit.  It is more likely to be
representative of the true stellar continuum than a smoothed spectrum.

\citet{she09} pointed out the importance of determining the continuum
accurately when measuring weak lines in medium-resolution spectra.
They refined their continuum determinations by iteratively fitting a
high-order spline to the quotient of the observed spectrum and the
best-fitting synthetic spectrum.  We adopted this procedure as well.
As part of the iterative process described in
Sec.~\ref{sec:iterations}, we fit a B-spline with a breakpoint spacing
of 50~\AA\ to the observed spectrum divided by the best-fitting
synthetic spectrum.  We divided the observed spectrum by this spline
before the next iteration of abundance measurement.


\section{Abundance Measurements}
\label{sec:measure}

The following section details some improvements on the abundance
measurement techniques of KGS08.  Aspects of the technique not
mentioned here were unchanged from the technique of KGS08.  In
summary, each observed spectrum was compared to a large grid of
synthetic spectra.  The atmospheric abundances were adopted from the
synthetic spectrum with the lowest $\chi^2$.

A major improvement was our measurement of four individual elemental
abundances in addition to Fe: Mg, Si, Ca, and Ti.  We chose these
elements because they are important in characterizing the star
formation history of a stellar population and because a significant
number of lines represent each of them in the DEIMOS spectral range.

\subsection{Model Atmospheres}

Like KGS08, we built synthetic spectra based on ATLAS9 model
atmospheres \citep{kur93} with no convective overshooting
\citep{cas97}.  KGS08 chose to allow the atmospheres to have $\mathafe
= +0.4$ or $\mathafe = 0.0$.  This choice allowed them to use the
large grid of ATLAS9 model atmospheres computed with new opacity
distribution functions \citep{cas04}.  However, we found that
best-fitting model spectra computed by KGS08 tended to cluster around
$\mathafe = +0.2$ due to the discontinuity in $\chi^2$ caused by the
abrupt switch between alpha-enhanced and solar-scaled models.

\begin{deluxetable}{lccc}
\tablewidth{0pt}
\tablecolumns{4}
\tablecaption{New Grid of ATLAS9 Model Atmospheres\label{tab:atlas}}
\tablehead{\colhead{Parameter} & \colhead{Minimum Value} & \colhead{Maximum Value} & \colhead{Step}}
\startdata
\teff~(K)           & 3500 & 5600 & 100 \\
                    & 5600 & 8000 & 200 \\
\logg~(cm~s$^{-2}$) & 0.0 ($\mathteff < 7000$~K)   &  5.0   &  0.5  \\
                    & 0.5 ($\mathteff \ge 7000$~K)   &  5.0   &  0.5  \\
{[A/H]}             & $-4.0$ & $0.0$  & $0.5$ \\
\afe                & $-0.8$ & $+1.2$ & $0.1$ \\
\enddata
\end{deluxetable}

To avoid this discontinuity, we recomputed ATLAS9 model atmospheres on
the grid summarized in Table~\ref{tab:atlas}.  The new grid required
recomputing new opacity distribution functions (ODFs), for which we
used the DFSYNTHE code \citep{cas05}.  Unlike the grid of
\citet{cas04}, we adopted the solar composition of \citet{and89},
except for Fe, for which we followed \citet[][see the note in
  Table~\ref{tab:solarspec}]{sne92}.  One opacity distribution
function was computed for each of the 189 combinations of [A/H] and
\afe\ specified in Table~\ref{tab:atlas}.  The abundances of all the
elements except H and He were augmented by [A/H].  Additionally, the
abundances of O, Ne, Mg, Si, Ar, Ca, and Ti were augmented by \afe.
These ODFs were used to compute one ATLAS9 model atmosphere for each
grid point in Table~\ref{tab:atlas} and for two values of
microturbulent velocity, for a total of 139104 model atmospheres.

\subsection{Microturbulent Velocity}
\label{sec:vt}

In order to reduce the number of parameters required to determine a
stellar abundance, KGS08 assumed that the microturbulent velocity
($\xi$) of the stellar atmosphere was tied to the surface gravity
(\logg).  They chose to fit a line to the spectroscopically measured
$\xi$ and \logg\ of the giant stars in \citeauthor{ful00}'s
(\citeyear{ful00}) sample:

\begin{equation}
\xi~(\mathrm{km~s}^{-1}) = 2.70 - 0.51\,\log g
\end{equation}

We also adopted a relation between $\xi$ and \logg, but we
re-determined this relation from the GC red giant sample of KGS08
combined with \citeauthor{kir09}'s (\citeyear{kir09}) compilation of
high-resolution spectroscopic measurements from the literature
\citep[][and references from
  KGS08]{fre09,gei05,joh02,lai07,she01,she03}.  The best-fit line
between the spectroscopically measured $\xi$ and \logg\ is

\begin{equation}
\label{eq:vtlogg} \xi~(\mathrm{km~s}^{-1}) = (2.13 \pm 0.05) - (0.23 \pm 0.03)\,\log g \: .
\end{equation}

\noindent
corresponding roughly to a 0.0--0.5~km~s$^{-1}$ decrease in $\xi$,
depending on \logg.  In the generation of the grid of synthetic
stellar spectra described in Sec.~\ref{sec:generation}, $\xi$ was not
a free parameter, but was fixed to \logg\ via Eq.~\ref{eq:vtlogg}.

In general, a decrease in $\xi$ increases the measurement of \feh.
Therefore, this change tended to increase the derived values of \feh.
A typical change in \feh\ was $\la +0.05$~dex.  This change would be
more severe in an HRS analysis based on equivalent widths (EWs).  In
our $\chi^2$ minimization, the abundance measurement was most
sensitive to lines with large $d({\rm EW})/d\mathfeh$.  Such lines are
the weak, unsaturated transitions whose strength does not depend on
$\xi$.  The DEIMOS spectra contain enough of these weak lines that
$\xi$ did not play a large role in the abundance determination.

\subsection{Line List}
\label{sec:linelist}

We compared the \ion{Fe}{1} oscillator strengths ($\log gf$) in the
KGS08 line list to values measured in the laboratory \citep{fuh06}.
Most of the KGS08 oscillator strengths were stronger than the
laboratory measurements.  The average offset was 0.13~dex.  Because
KGS08 calibrated their line list to the solar spectrum, we interpreted
this offset as a systematic error in the solar model atmosphere, solar
spectral synthesis, and/or solar composition.  Accepting the
laboratory-measured values as more accurate than the solar
calibration, we replaced \ion{Fe}{1} oscillator strengths with
\citeauthor{fuh06} where available, and we subtracted 0.13~dex from
$\log gf$ for all other \ion{Fe}{1} transitions in the KGS08 line
list.  All other data remained unchanged.

Decreasing the oscillator strengths requires a larger \feh\ to match
the observed spectrum.  The amount of change in \feh\ depends on the
atmospheric parameters as well as the saturation of the measured Fe
lines.  From comparison of results with the old and new line lists, we
estimate a typical change in \feh\ to be $\sim +0.1$~dex.

\subsection{Generation of Synthetic Spectra}
\label{sec:generation}

The spectra were synthesized as described in KGS08.  Specifically, the
current version of the local thermodynamic equilibrium (LTE) spectrum
synthesis software MOOG \citep{sne73} generated one spectrum for each
point on the grid.  The spectral grid was more finely spaced in
\feh\ than the model atmosphere grid.  The spacing is 0.1~dex for each
of \feh\ and \afe, yielding a total of 316848 synthetic spectra.

\begin{deluxetable}{lr|lr}
\tablewidth{0pt}
\tablecolumns{4}
\tablecaption{Adopted Solar Composition\label{tab:solarspec}}
\tablehead{\colhead{Element} & \colhead{$12 + \log \epsilon$} & \colhead{Element} & \colhead{$12 + \log \epsilon$}}
\startdata
Mg & 7.58 & Ti & 4.99 \\
Ca & 6.36 & Fe & 7.52 \\
Si & 7.55 &    &      \\
\enddata
\tablecomments{This composition is adopted from \protect\citet{and89},
  except for Fe.  For justification of the adopted Fe solar abundance,
  see \citet{sne92}.  The abundance of an element X is defined as its
  number density relative to hydrogen: $12+ \log \epsilon_{\rm{X}} =
  12 + \log (n_{\rm{X}}) - \log (n_{\rm{H}})$.}
\end{deluxetable}

The solar composition used in the generation of the synthetic spectra
was identical to the solar composition used in the computation of the
model atmospheres.  Table~\ref{tab:solarspec} lists the adopted solar
abundances for the five elements for which we measure abundances in
Sculptor stars.

\subsection{Effective Temperatures and Surface Gravities}
\label{sec:tefflogg}

Different spectroscopic studies of chemical abundances rely on
different sources of information for determining the effective
temperature (\teff) and surface gravity (\logg) of the stellar
atmosphere.  KGS08 consulted Yonsei-Yale model isochrones
\citep{dem04} to determine the temperature and gravity that correspond
to a dereddened color and an extinction-corrected absolute magnitude.
They also considered Victoria-Regina \citep{van06} and Padova
\citep{gir02} model isochrones, as well as an empirical
color-temperature relation \citep{ram05}.

The Fe lines accessible in DEIMOS spectra span a large range of
excitation potential.  Together, these different lines provide a
constraint on \teff.  KGS08 (their Sec.~5.1) showed that---without any
photometric information---the synthesis analysis of medium-resolution
spectra of GC stars yielded values of \teff\ very close to values
previously measured from HRS.  Therefore, we chose to measure
\teff\ from photometry and spectroscopy simultaneously.

To begin, we converted extinction-corrected \citep{sch98} Washington
$M$ and $T_2$ magnitudes to Cousins $V_{\rm C}$ and $I_{\rm C}$
magnitudes \citep{maj00}.  With these magnitudes, we computed
\teff\ from the Yonsei-Yale, Victoria-Regina, and Padova model
isochrones, as well as the \citet{ram05} empirical color-based \teff.
For each measurement, we estimated the effect of photometric error by
measuring the standard deviation of \teff\ determined from 1000 Monte
Carlo realizations of $V_{\rm C}$ and $I_{\rm C}$.  In each
realization, $V_{\rm C}$ and $I_{\rm C}$ were chosen from a normal
distribution with a mean of the measured, extinction-corrected
magnitude and a standard deviation of the photometric error.  We call
this error $\delta T_{{\rm eff,}i}$, where $i$ represents each of the
four photometric methods of determining \teff.  In order to arrive at
a single photometric \teff, we averaged the four $T_{{\rm eff,}i}$
together with appropriate error weighting.  We also estimated the
random and systematic components of error.  In summary,

\begin{eqnarray}
\label{eq:teffphot} \overline{T_{\rm{eff}}} &=& \frac{\sum_i T_{{\rm eff,}i} \delta T_{{\rm eff,}i}^{-2}}{\sum_i \delta T_{{\rm eff,}i}^{-2}} \\
\delta_{\rm{rand}} T_{\rm{eff}} &=& \frac{\sum_i \delta T_{{\rm eff,}i}^{-1}}{\sum_i \delta T_{{\rm eff,}i}^{-2}} \\
\delta_{\rm{sys}} T_{\rm{eff}} &=& \sqrt{\frac{\sum_i \delta T_{{\rm eff,}i}^{-2} \sum_i \delta T_{{\rm eff,}i}^{-2} \left(T_{{\rm eff,}i} - \overline{T_{\rm{eff}}}\right)^2}{1 - \left(\sum_i \delta T_{{\rm eff,}i}^{-2}\right)^2 \sum_i \delta T_{{\rm eff,}i}^{-4}}} \\
\label{eq:tefferr} \delta_{\rm{total}} T_{\rm{eff}} &=& \sqrt{(\delta_{\rm{rand}} T_{\rm{eff}})^2 + (\delta _{\rm{sys}}T_{\rm{eff}})^2}
\end{eqnarray}

For the stars in this data set, the median random, systematic, and
total errors on \teff\ were \sclmedtefferrrand~K, \sclmedtefferrsys~K,
and \sclmedtefferrtot~K respectively.  The somewhat large errors on
the photometric temperatures indicated that the spectra may help
constrain \teff.  Therefore, Eq.~\ref{eq:teffphot} does not show the
final temperature used in the abundance determination.
Section~\ref{sec:iterations} describes the iterative process for
determining \teff\ and elemental abundances from spectroscopy.

We followed a similar procedure for determining
\logg\ photometrically, except that we used only the three model
isochrones and not any empirical calibration.  The error on the true
distance modulus \citep[$19.67 \pm 0.12$,][]{pie08} was included in
the Monte Carlo determination of the error on \logg.  The median
random, systematic, and total errors on \logg\ were
\sclmedloggerrrand, \sclmedloggerrsys, and \sclmedloggerrtot.  These
errors are very small, and the medium-resolution, red spectra have
little power to help constrain \logg\ because there are so few ionized
lines visible.  Therefore, we assumed the photometric value of
\logg\ for the abundance analysis.

\subsection{Wavelength Masks}
\label{sec:masks}

The procedure described in the next section consisted of separately
measuring the abundances of five elements: Mg, Si, Ca, Ti, and Fe.
The procedure relied on finding the synthetic spectrum that best
matched an observed spectrum.  In order to make this matching most
sensitive to a particular element, we masked all spectral regions that
were not significantly affected by abundance changes of that element.

To make the wavelength masks, we began with a base spectrum that
represented the solar composition in which the abundances of all the
metals were scaled down by 1.5~dex ($\rm{[A/H]} = -1.5$).  The
temperature and gravity of the synthetic star were $\mathteff =
4000$~K and $\mathlogg = 1.0$.  Then, we created two pairs of spectra
for each of the five elements.  In one spectrum, the abundance of the
element was enhanced by 0.3~dex, and in the other, depleted by
0.3~dex.  Spectral regions where the flux difference between these two
spectra exceeds 0.5\% were used in the abundance determination of that
element.  This small threshold assured that weak lines, which
experience large fractional changes in EW as \feh\ changes, were
included in the analysis.  We repeated this procedure for spectra with
$\mathteff = 5000$~K, 6000~K, 7000~K, and 8000~K.  Additional spectral
regions that passed the 0.5\% flux difference criterion were also
included in the abundance determination of that element.  All other
wavelengths were masked.

\begin{figure}[t!]
\includegraphics[width=\linewidth]{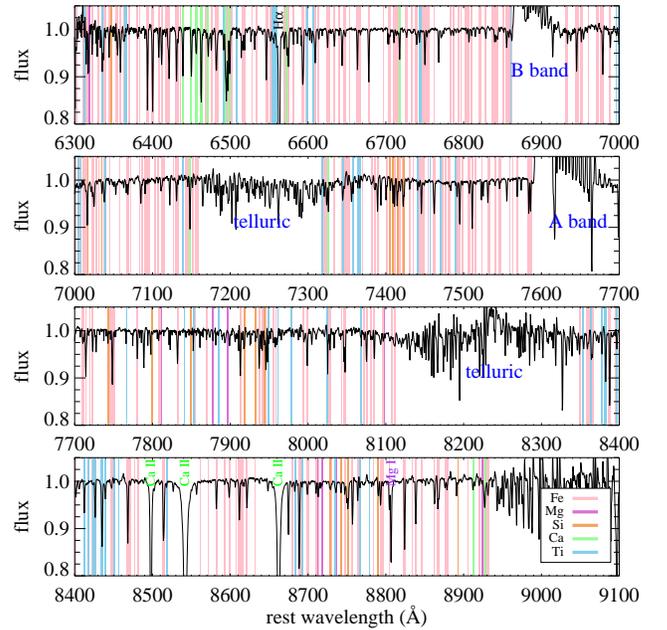}
\caption{A coaddition of all Sculptor stars with the continuum
  normalized to unity.  The high SNR provided by the coaddition makes
  stellar absorption lines readily apparent.  The colored regions show
  the wavelength masks used in the determination of the abundance of
  each element.  Regions susceptible to telluric absorption are
  labeled with blue text.  Because large residuals from the telluric
  absorption correction remain, we eliminate these regions from the
  abundance analysis.  Some stellar features excluded from the
  abundances measurement are also labeled.\label{fig:coadd}}
\end{figure}

The result was one wavelength mask for each of Mg, Si, Ca, Ti, and Fe,
shown in Fig.~\ref{fig:coadd}.  We also created one ``$\alpha$'' mask
as the intersection of the Mg, Si, Ca, and Ti masks.  The $\alpha$
element regions do not overlap with each other, but the $\alpha$
element regions do overlap with the Fe regions.  The most severe case
is the Ca mask, where $\sim 35\%$ of the pixels are shared with the Fe
mask.  However, the overlap did not introduce interdependence in the
abundance measurements.  The $\alpha$ element abundances were held
fixed while \feh\ was measured, and the Fe abundance was held fixed
while \afe\ was measured.  The measurements of \feh\ and \afe\ were
performed iteratively (see the next subsection).  We tested the
independence of the measurements by removing all overlapping pixels
from consideration.  Abundance measurements changed on average by only
0.01~dex.

\subsection{Measuring Atmospheric Parameters and Elemental Abundances}
\label{sec:iterations}

A Levenberg-Marquardt algorithm \citep[the IDL routine MPFIT, written
by][]{mark09} found the best-fitting synthetic spectrum in ten
iterative steps.  In each step, the $\chi^2$ was computed between an
observed spectrum and a synthetic spectrum degraded to match the
resolution of the observed spectrum.  First, we interpolated the
synthetic spectrum onto the same wavelength array as the observed
spectrum.  Then, we smoothed the synthetic spectrum through a Gaussian
filter whose width was the observed spectrum's measured resolution as
a function of wavelength.

\begin{enumerate}
\item \teff\ and \feh, first pass: An observed spectrum was compared
  to a synthetic spectrum with \teff\ and \logg\ determined as
  described in Sec.~\ref{sec:tefflogg} and \feh\ determined from
  Yonsei-Yale isochrones.  For this iteration, \afe\ was fixed at 0.0
  (solar), and only spectral regions most susceptible to Fe absorption
  (Sec.~\ref{sec:masks}) were considered.  The two quantities \teff\
  and \feh\ were varied, and the algorithm found the best-fitting
  synthetic spectrum by minimizing $\chi^2$.  We sampled the parameter
  space between grid points by linearly interpolating the synthetic
  spectra at the neighboring grid points.  \teff\ was also loosely
  constrained by photometry.  As the spectrum caused \teff\ to stray
  from the photometric values, $\chi^2$ increased, and it increased
  more sharply for smaller photometric errors (as calculated in
  Eq.~\ref{eq:tefferr}).  Therefore, both photometry and spectroscopy
  determined \teff.  Photometry alone determined \logg.

\item \afe, first pass: For this iteration, \teff, \logg, and \feh\
  were fixed.  Only \afe\ was allowed to vary.  In the model stellar
  atmosphere, the abundances of the $\alpha$ elements with respect to
  Fe varied together.  Only the spectral regions susceptible to
  absorption by Mg, Si, Ca, or Ti were considered.

\item Continuum refinement: The continuum-divided, observed spectrum
  was divided by the synthetic spectrum with the parameters determined
  in steps 1 and 2.  The result approximated a flat noise spectrum.
  To better determine the continuum, we fit a B-spline with a
  breakpoint spacing of 50~\AA\ to the residual spectrum.  We divided
  the observed spectrum by the spline fit.

\item \feh, second pass: We repeated step 1 with the revised spectrum,
but \teff\ was held fixed at the previously determined value.

\item{[Mg/Fe]: We repeated step 2.  However, only Mg spectral lines
  were considered in the abundance measurement.}

\item{[Si/Fe]: We repeated step 5 for Si instead of Mg.}

\item{[Ca/Fe]: We repeated step 5 for Ca instead of Mg.}

\item{[Ti/Fe]: We repeated step 5 for Ti instead of Mg.}

\item \afe, second pass: We repeated step 2 for all of the $\alpha$
  elements instead of just Mg.  This step was simply a different way to
  average the $\alpha$ element abundances than combining the
  individual measurements of [Mg/Fe], [Si/Fe], [Ca/Fe], and [Ti/Fe].

\item \feh, third pass: The value of \afe\ affected the measurement of
  \feh\ because \afe\ can affect the structure of the stellar
  atmosphere.  Specifically, the greater availability of electron
  donors with an increased \afe\ ratio allows for a higher density of
  H$^{-}$ ions.  The subsequent increase in continuous opacity
  decreases the strength of Fe and other non-$\alpha$ element lines.
  With \afe\ fixed at the value determined in step 9, we re-measured
  \feh.  Typically, \feh\ changed from the value determined in step 1
  by much less than 0.1~dex.
\end{enumerate}


\subsection{Correction to [Fe/H]}
\label{sec:correction}

In comparing our MRS measurements of \feh\ to HRS measurements of the
same stars (see the appendix), we noticed that our measurements of
metal-poor stars were consistently $\sim 0.15$~dex lower.  The same
pattern is also visible in the \citet{kir08a} GC measurements (see
their Figs.~6, 7, 10, and 11).

We have thoroughly examined possible sources of this difference of
scale.  The changes to the microturbulent velocity relation
(Sec.~\ref{sec:vt}) and the line list (Sec.~\ref{sec:linelist}) were
intended to yield a more accurate and standardized estimation of \feh,
but the offset still remained.  Restricting the analysis to narrow
spectral regions did not reveal any systematic trend of \feh\ with
wavelength.

A possible explanation for this offset is overionization
\citep{the99}.  Ultraviolet radiation in stellar atmospheres can
ionize Fe more than would be expected in LTE.  Therefore, the
abundance of \ion{Fe}{1} would seem to be lower than the abundance of
\ion{Fe}{2} in an LTE analysis.  \ion{Fe}{2} does not suffer from this
effect.  However, the effect is smaller at higher \feh, and we do not
observe a trend with metallicity for the offset of our values relative
to HRS studies.

In order to standardize our measurements with previous HRS studies, we
added 0.15~dex to all of our measurements of \feh.  This offset and
the microturbulent velocity-surface gravity relation are the only ways
in which previous HRS studies inform our measurements.  Furthermore,
this offset is not intended to change the standardization of our
abundances.  All of the abundance in this article, including those
from other studies, are given relative to the solar abundances quoted
in Table~\ref{tab:solarspec}.

\subsection{Error Estimation}
\label{sec:error}

\begin{deluxetable}{lr|lr}
\tablewidth{0pt}
\tablecolumns{4}
\tablecaption{Systematic Abundance Errors\label{tab:syserr}}
\tablehead{\colhead{Element Ratio} & \colhead{$\delta_{\rm sys}$} & \colhead{Element Ratio} & \colhead{$\delta_{\rm sys}$}}
\startdata
\protect[Fe/H]  & \fehsyserr  & [Ca/Fe] & \cafesyserr \\
\protect[Mg/Fe] & \mgfesyserr & [Ti/Fe] & \tifesyserr \\
\protect[Si/Fe] & \sifesyserr &         &             \\
\enddata
\end{deluxetable}

We repeated the error estimation procedure described by KGS08 (their
Sec.~6) by repeating their abundance analysis on GC stars with the
above modifications.  We no longer found a convincing trend of
$\delta\mathfeh$ with \feh.  Instead, we estimate the total error on
\feh\ by adding a systematic error in quadrature with the
SNR-dependent uncertainty of the synthetic spectral fit.  The
magnitude of $\delta_{\rm{sys}}\mathfeh = \fehsyserr$ was the value
required to force HRS and MRS \feh\ estimates of the same GC stars to
agree at the 1$\sigma$ level.  We also estimated systematic errors for
each of [Mg/Fe], [Si/Fe], [Ca/Fe], and [Ti/Fe] in the same manner as
for \feh.  These are listed in Table~\ref{tab:syserr}.


\section{Results}
\label{sec:abund}

In this section, we discuss the interpretation of the abundance
measurements in Sculptor, all of which are presented in
Table~\ref{tab:data} on the last page of this manuscript.

\subsection{Metallicity Distribution}

The metallicity distribution function (MDF) of a dwarf galaxy can
reveal much about its star formation history.  In chemical evolution
models of dwarf galaxies \cite[e.g.,][]{lan04,mar06,mar08}, the
duration of star formation affects the shape of the MDF.  The MDF also
has implications for the formation of the MW.  If the MW halo was
built from dSphs \citep{sea78,whi78}, then it is important to find
dSph counterparts to halo field stars at all metallicities, as pointed
out by \citet[][hereafter H06]{hel06}.

\begin{figure}[t!]
\includegraphics[width=\linewidth]{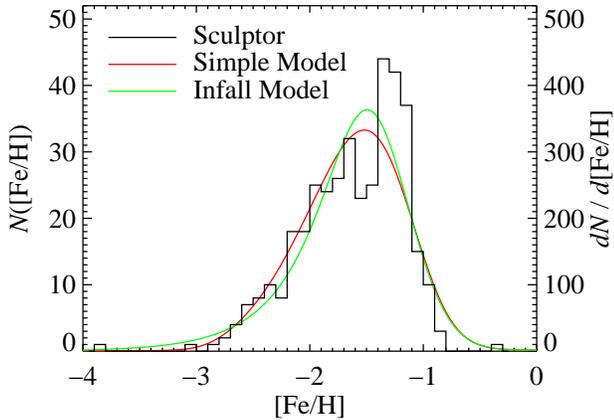}
\caption{The metallicity distribution in Sculptor.  The red curve is
  the maximum likelihood fit to a galactic chemical evolution model
  with pre-enrichment (Eq.~\ref{eq:gce}), and the green curve is the
  maximum likelihood fit to a model of star formation in the presence
  of infalling, zero-metallicity gas (Eq.~\ref{eq:infallgce}).  The
  long, metal-poor tail is typical for systems with non-instantaneous
  star formation.\label{fig:feh_hist}}
\end{figure}

Figure~\ref{fig:feh_hist} shows the MDF of Sculptor.  The shape of the
MDF is highly asymmetric, with a long, metal-poor tail \citep[as
predicted by][]{sal09}.  The inverse-variance weighted mean is
$\langle\mathfeh\rangle = \sclfehmean$ with a standard deviation of
$\sclfehsigma$.  The median is $\sclfehmedian$ with a median absolute
deviation of $\sclfehmad$ and an interquartile range of $\sclfehiqr$.

\begin{figure}[t!]
\includegraphics[width=\linewidth]{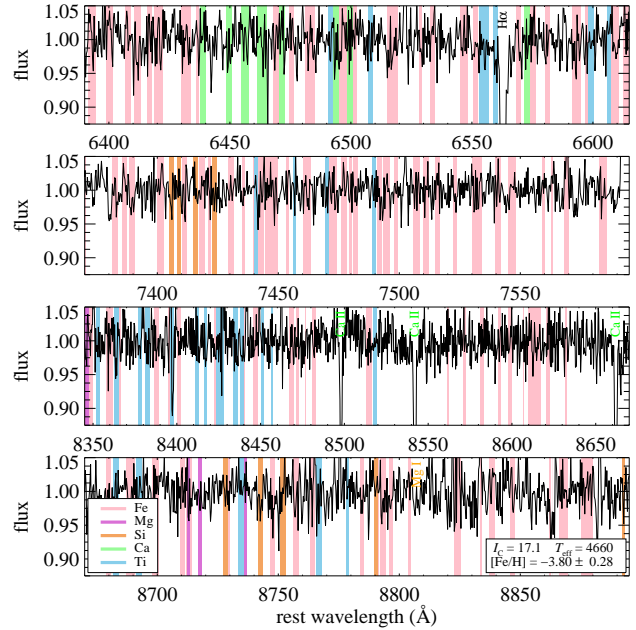}
\caption{Regions of the DEIMOS spectrum of the extremely metal-poor
  star S1020549, which has $\mathfeh = \sclempfehtwo \pm
  \sclempfeherrtwo$.  The spectrum appears particularly noisy because
  the $y$-axis range is small.  Some Fe absorption lines are barely
  detectable, but all together, they contain enough signal to make a
  quantitative measurement of \feh.  The shading corresponds to the
  same spectral region shown in Fig.~\ref{fig:coadd}.  Frebel, Kirby,
  \& Simon (in preparation) will present a high-resolution spectrum of
  this star, which confirms the extremely low
  metallicity.\label{fig:1020549}}
\end{figure}

The MDF boasts an exceptionally metal-poor star, S1020549.  The
metallicity is $\mathfeh = \sclempfehtwo \pm \sclempfeherrtwo$.
Figure~\ref{fig:1020549} shows how weak the Fe absorption lines are in
this star.  Frebel, Kirby, \& Simon (in preparation) have confirmed
this extremely low metallicity with a high-resolution spectrum.

Sculptor is now the most luminous dSph in which an extremely
metal-poor (EMP, $\mathfeh < -3$) star has been detected.
[\citet{kir08b} discovered 15 EMP stars across eight ultra-faint dwarf
galaxies, and \citet{coh09} discovered one EMP star in the
\object[NAME DRACO DSPH GALAXY]{Draco dSph}.]  Stars more metal-poor
than S1020549 are known to exist only in the field of the Milky Way
field.  This discovery hints that dSph galaxies like Sculptor may have
contributed to the formation of the metal-poor component of the halo.
We discuss Sculptor's link to the halo further in
Sec.~\ref{sec:halomdf}.

The $MT_2D$ photometric selection of spectroscopic targets may have
introduced a tiny \feh\ bias.  Figure~\ref{fig:cmd} shows that the RGB
is sharply defined in Sculptor.  Because the number density of stars
redward and blueward of the RGB is much lower than the number density
on the RGB, the number of very young or very metal-poor stars
(blueward) or very metal-rich stars (redward) missed by photometric
pre-selection must be negligible.  Furthermore, the hard color cut (as
opposed to one that depends on $M-D$ color) was $0.6 < (M - T_2)_0 <
2.2$.  The CMD gives no reason to suspect Sculptor RGB members outside
of these limits, but it is possible that some extremely blue Sculptor
members have been excluded.

\subsubsection{Possible Explanation of the Discrepancy with Previous Results}

Our measured MDF and our detection of EMP stars in Sculptor are at
odds with the findings of H06.  Whereas our MDF peaks at $\mathfeh
\sim -1.3$, theirs peaks at $\mathfeh \sim -1.8$.  Furthermore, our
observed MDF is much more asymmetric than that of H06, which may even
be slightly asymmetric in the opposite sense (a longer metal-{\it
rich} tail).  The greater symmetry would indicate a less extended star
formation history or early infall of a large amount of gas
\citep{pra03}.

\begin{figure}[t!]
\includegraphics[width=\linewidth]{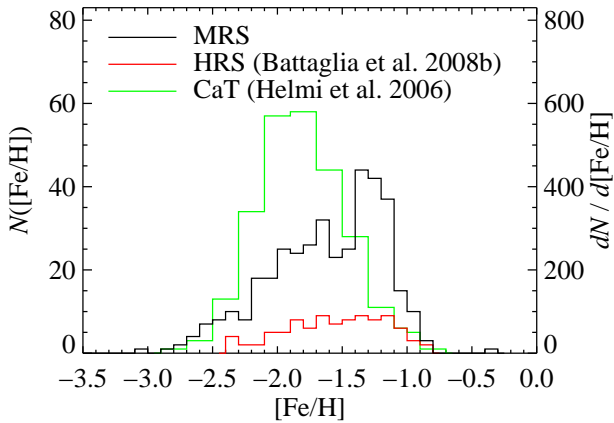}
\caption{Sculptor's metallicity distribution as observed in this study
  (MRS, {\it black}) and by B08b (HRS, {\it red}), which is a subset
  of the MDF observed by H06 (Ca triplet, {\it green}).  The CaT-based
  MDF is more metal-poor probably because the sample of H06 is more
  spatially extended than the other two
  samples.\label{fig:feh_hist_bat08}}
\end{figure}

\citet[][hereafter B08b]{bat08b} observed a subset of the H06 stars at
high resolution.  The MDFs from the two studies have noticeably
different shapes.  Figure~\ref{fig:feh_hist_bat08} shows that the HRS
MDF peaks at $\mathfeh \sim -1.3$, which is also the peak that we
observe.  The mean and standard deviation of their MDF are
$\sclmdfbatmean$ and $\sclmdfbatsigma$.  However, the MDF of H06 peaks
at $\mathfeh \sim -1.8$, and the mean and standard deviation are
$\sclmdfhelmean$ and $\sclmdfhelsigma$.  The overlapping stars between
the samples of B08b and H06 agree very well.

The most likely explanation for the different MDFs is the different
spatial sampling of the three studies.  Sculptor has a steep radial
metallicity gradient \citep[][also see
  Sec.~\ref{sec:gradients}]{tol04,wes06,wal09}.  The stars in the
center of Sculptor are more metal-rich than stars far from the center.
H06 sampled stars out to the tidal radius \citep[$r_t =
  76.5$~arcmin,][]{mat98}, but we and B08b sampled stars only out to
about 11~arcmin.  As a result, the mean metallicity of the H06 CaT
sample is lower than our MRS sample and the B08b HRS sample.  In the
next subsection, we address the chemical evolution of Sculptor based
on its MDF.  Our conclusions are based only on stars within the
central 11~arcmin.

\subsubsection{Quantifying Chemical Evolution in Sculptor}

In chemical evolution models, extended star formation produces a long,
metal-poor tail.  \citet{pra08} described the shape of the
differential metallicity distribution derived from a ``simple model''
of galactic chemical evolution.  Expressed in terms of \feh\ instead
of metal fraction $Z$, the predicted distribution is

\begin{equation}
\frac{dN}{d\mathfeh} = A\left(10^{\mathfeh} - 10^{\mathfeh_i}\right) \exp \left(-\frac{10^{\mathfeh}}{p}\right)\label{eq:gce}
\end{equation}

\noindent where $p$ is the effective yield in units of the solar metal
fraction ($Z_\sun$) and $\mathfeh_i$ is the initial gas metallicity.
An initial metallicity is needed to resolve the Galactic G dwarf
problem \citep{van62,sch63}.  $A$ is a normalization that depends on
$p$, $\mathfeh_i$, the final metallicity $\mathfeh_f$, and the number
of stars in the sample $N$:

\begin{equation}
A = \frac{(N \ln 10)/p}{\exp\left(-\frac{10^{\mathfeh_i}}{p}\right) -
  \exp\left(-\frac{10^{\mathfeh_f}}{p}\right)}\label{eq:norm}
\end{equation}

The red curve in Figure~\ref{fig:feh_hist} is the two-parameter,
maximum likelihood fit to Eq.~\ref{eq:gce}.  The likelihood $L_i$ that
star $i$ is drawn from the probability distribution defined by
Eq.~\ref{eq:gce} is the integral of the product of the error
distribution for the star and the probability distribution.  The total
likelihood $L = \prod_i L_i$.  The most likely $p$ and $\mathfeh_0$
are the values that maximize $L$.  For display, the curve has been
convolved with an error distribution, which is a composite of $N$ unit
Gaussians.  $N$ is the total number of stars in the observed
distribution, and the width of the $i^{\rm th}$ Gaussian is the
estimated total \feh\ error on the $i^{\rm th}$ star.  This
convolution approximates the effect of measurement error on the model
curve under the assumption that the error on \feh\ does not depend on
\feh.  This assumption seems to be valid because our estimates of
$\delta\mathfeh$ do not show a trend with \feh.

The most likely yield---largely determined by the \feh\ at the peak of
the MDF---is $p = \sclsimpleyield Z_\sun$.  [From the MDF of H06,
  \citet{pra08} calculated $p = 0.016 Z_\sun$.]  We also measure
$\mathfeh_0 = \sclfehinitial$.  H06 also measured $\mathfeh_0 = -2.90
\pm 0.21$ for Sculptor, even though they included stars out to the
tidal radius, which are more metal-poor on average than the centrally
concentrated stars in our sample.  (Instead of finding the maximum
likelihood model, they performed a least-squares fit to the cumulative
metallicity distribution without accounting for experimental
uncertainty.  In general, observational errors exaggerate the extrema
of the metallicity distribution, and the least-squares fit converges
on a lower $\mathfeh_0$ than the maximum likelihood fit.)  One
explanation that they proposed for this non-zero initial metallicity
was pre-enrichment of the interstellar gas that formed the first
stars.  Pre-enrichment could result from a relatively late epoch of
formation for Sculptor, after the supernova (SN) ejecta from other
galaxies enriched the intergalactic medium from which Sculptor formed.
However, our observation of a star at $\mathfeh = \sclempfehtwo$ is
inconsistent with pre-enrichment at the level of $\mathfeh_0 = -2.9$.

\citet{pra08} instead interpreted the apparent dearth of EMP stars as
an indication of early gas infall \citep{pra03}, wherein star
formation begins from a small amount of gas while the majority of gas
that will eventually form dSph stars is still falling in.  In order to
test this alternative to pre-enrichment, we have also fit an Infall
Model, the ``Best Accretion Model'' of \citet[][also see
  \citeauthor{pag97} \citeyear{pag97}]{lyn75}.  It is one of the
models which accounts for a time-decaying gas infall that has an
analytic solution.  The model assumes that the gas mass $g$ in units
of the initial mass is related quadratically to the stellar mass $s$
in units of the initial mass:

\begin{equation}
g(s) = \left(1 - \frac{s}{M}\right)\left(1 + s - \frac{s}{M}\right)\label{eq:g}
\end{equation}

\noindent
where $M$ is a parameter greater than 1.  When $M=1$, Eq.~\ref{eq:g}
reduces to $g = 1 - s$, which describes the Closed Box Model.
Otherwise, $M$ monotonically increases with the amount of gas infall
and with the departure from the Simple Model.  Following \citet{lyn75}
and \citet{pag97}, we assume that the initial and infalling gas
metallicity is zero.  The differential metallicity distribution is
described by two equations.

\begin{eqnarray}
\mathfeh(s) &=& \log \Big\{p \left(\frac{M}{1 + s -
  \frac{s}{M}}\right)^2 \times \label{eq:s} \\
\nonumber & & \left[\ln \frac{1}{1 - \frac{s}{M}} - \frac{s}{M} \left(1 - \frac{1}{M}\right)\right]\Big\} \\
\frac{dN}{d\mathfeh} &=& A \frac{10^{\mathfeh}}{p} \times \label{eq:infallgce} \\
\nonumber & & \frac{1 + s\left(1 - \frac{1}{M}\right)}{\left(1 - \frac{s}{M}\right)^{-1} - 2 \left(1 - \frac{1}{M}\right) \times 10^{\mathfeh/p}}
\end{eqnarray}

\noindent
Equation~\ref{eq:s} is transcendental, and it must be solved for $s$
numerically.  Equation~\ref{eq:infallgce} decouples the peak of the
MDF from the yield $p$.  As $M$ increases, the MDF peak decreases
independently of $p$.

The green line in Fig.~\ref{fig:feh_hist} shows the most likely Infall
Model convolved with the error distribution as described above.  The
Infall Model has $M = \sclinfallm$, which is only a small departure
from the Simple Model.

Neither the Simple Model nor the Infall Model fits the data
particularly well.  Both models fail to reproduce the sharp peak at
$\mathfeh \sim -1.3$ and the steep metal-rich tail.  However, the
Infall Model does reproduce the metal-poor tail about as well as the
Simple Model.  Therefore, the Infall Model is a reasonable alternative
to pre-enrichment, and it allows the existence of the star at
$\mathfeh = \sclempfehtwo$.  In reality, a precise explanation of the
MDF will likely incorporate the radial metallicity gradients and
multiple, superposed populations.  It is tempting to conclude from
Fig.~\ref{fig:feh_hist} that Sculptor displays two metallicity
populations.  We have not attempted a two-component fit, but that
would seem to be a reasonable approach for future work, especially in
light of \citeauthor{tol04}'s (\citeyear{tol04}) report of two
distinct stellar populations in Sculptor.

Searches for the lowest metallicity stars in the MW halo have revealed
some exquisitely metal-poor stars \citep[e.g., $\mathfeh =
  -5.96$,][]{fre08}.  Such exotic stars have not yet been discovered
in any dSph.  However, if Sculptor was not pre-enriched, a large
enough sample of \feh\ measurements in Sculptor---and possibly other
dSphs---may reveal stars as metal-poor as the lowest metallicity stars
in the MW halo.

\subsubsection{Comparison to the Milky Way Halo MDF}
\label{sec:halomdf}

\begin{figure}[t!]
\includegraphics[width=\linewidth]{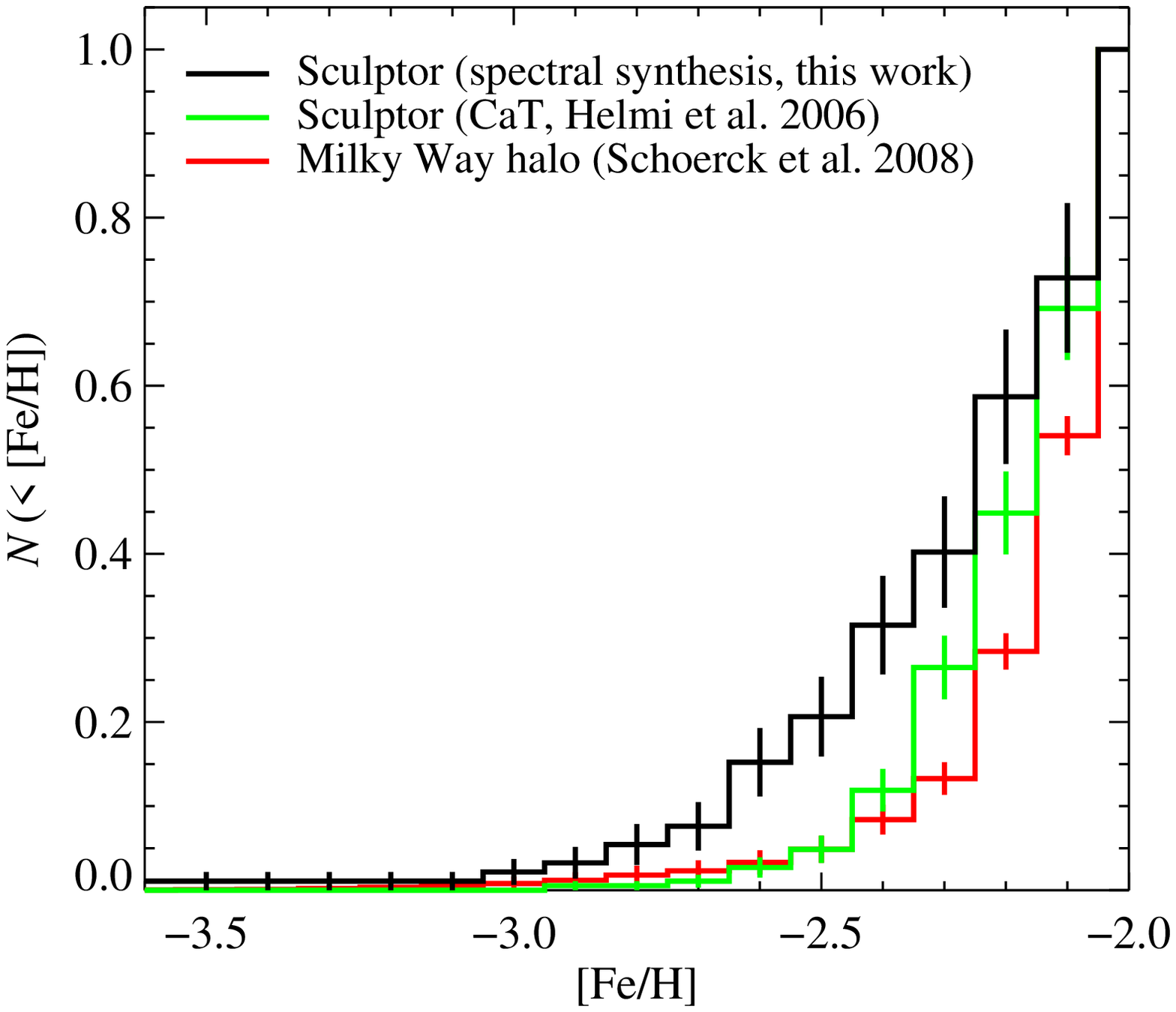}
\caption{The metal-poor tails of the MDFs in Sculptor ({\it black}) and
  Galactic halo field stars \citep[{\it red},][]{sch08} shown as
  cumulative distributions, all normalized to the number of stars with
  $\mathfeh < -2$.  The green line shows the MDF measured by the DART
  team \citep{hel06} with a calibration based on the Ca triplet.  The
  calibration may overpredict very low metallicities.  The
  synthesis-based metallicities ({\it black}, this work) are valid at
  lower \feh\ than the Ca triplet \feh.  Regardless, the halo has a
  steeper metal-poor tail than Sculptor in both representations.
  Galaxies such as Sculptor were probably not the dominant
  contributors to the halo.\label{fig:halo_compare}}
\end{figure}

\citet{sea78} and \citet{whi78} posited that the MW halo formed from
the accretion and dissolution of dwarf galaxies.  The dSphs that exist
today may be the survivors from the cannibalistic construction of the
Galactic halo.  \citet{hel06} suggested that at least some of the halo
field stars could not have come from counterparts to the surviving
dSphs because the halo field contained extremely metal-poor stars
whereas the dSphs do not.  However, \citet{sch08} showed that the
Hamburg/ESO Survey's halo MDF, after correction for selection bias,
actually looks remarkably like the MDFs of the dSphs Fornax,
\object[NAME UMi dSph]{Ursa Minor}, and Draco.  Furthermore,
\citet{kir08b} presented MRS evidence for a large fraction of EMP
stars in the ultra-faint dSph sample of \citet{sim07}, suggesting that
today's surviving dSphs contain stars that span the full range of
metallicities displayed by the Galactic field halo population.

We revisit the halo comparison with the present MDF for Sculptor.
Figure~\ref{fig:halo_compare} shows the metal-poor tail ($\mathfeh <
-2$) of the MRS synthesis-based Sculptor MDF presented here, the
CaT-based Sculptor MDF \citep{hel06}, and the MW halo MDF
\citep{sch08}.  As observed in the comparisons to other dSphs
presented by \citeauthor{sch08}, the halo seems to have a steeper
metal-poor tail than the CaT-based Sculptor MDF, despite the evidence
that CaT-based metallicities overpredict \feh\ at $\mathfeh \la -2.2$
\citep[e.g.,][]{koc08,nor08}.  The synthesis-based MDF does not rely
on empirical calibrations, and the technique has been shown to work at
least down to $\mathfeh = -3$ \citep{kir08b}.

This MDF shows that the halo has a much steeper metal-poor tail than
Sculptor.  This result is consistent with a merging scenario wherein
several dwarf galaxies significantly larger than Sculptor contributed
most of the stars to the halo field \citep[e.g.,][]{rob05,fon06}.  In
these models, the more luminous galaxies have higher mean
metallicities.  Galaxies with a Sculptor-like stellar mass are
minority contributors to the halo field star population.  Less
luminous galaxies are even more metal-poor \citep{kir08b}.  Therefore,
Sculptor conforms to the luminosity-metallicity relation for dSphs,
and the difference between Sculptor's MDF and the MW halo MDF does not
pose a problem for hierarchical assembly.

\subsection{Alpha Element Abundances}

\begin{figure}[t!]
\includegraphics[width=\linewidth]{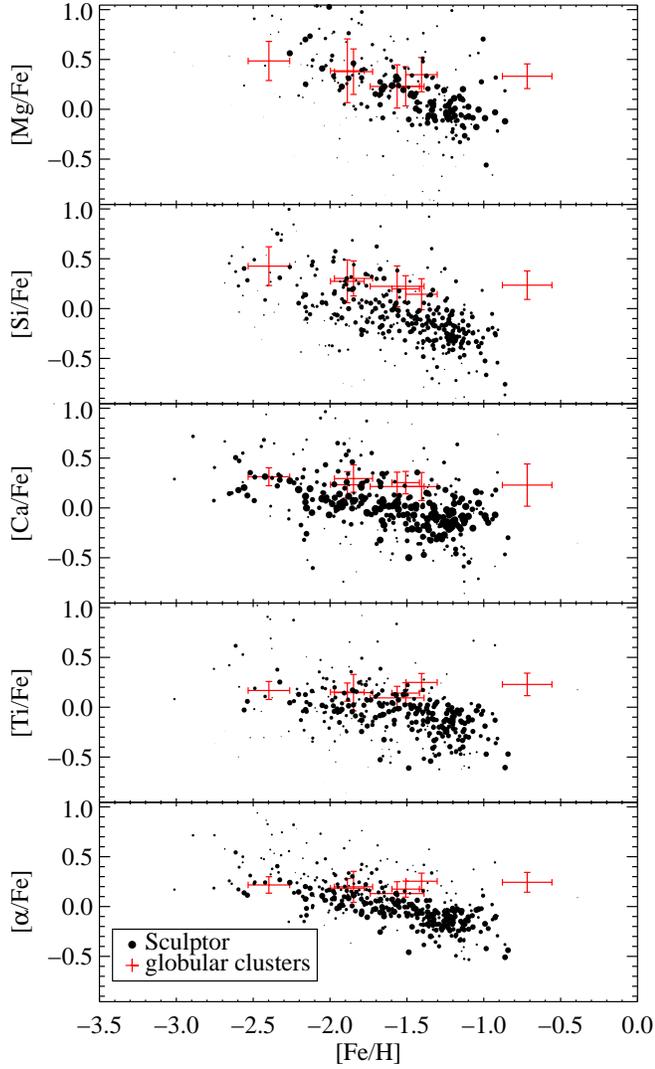}
\caption{Multi-element abundances in Sculptor ({\it black}).  The
  point sizes reflect the quadrature sum of the errors on \feh\ and
  [X/Fe], where larger points have smaller errors.  The bottom panel
  shows the average of the four elements shown in the other panels.
  For comparison, the red error bars show the means and standard
  deviations from the seven GCs of KGS08.  Because the Sculptor and GC
  abundances were measured in the same way, the comparison
  demonstrates that [X/Fe] declines with increasing \feh\ in Sculptor,
  but not the GCs.\label{fig:alphafe_feh_gc}}
\end{figure}

\begin{figure}[t!]
\includegraphics[width=\linewidth]{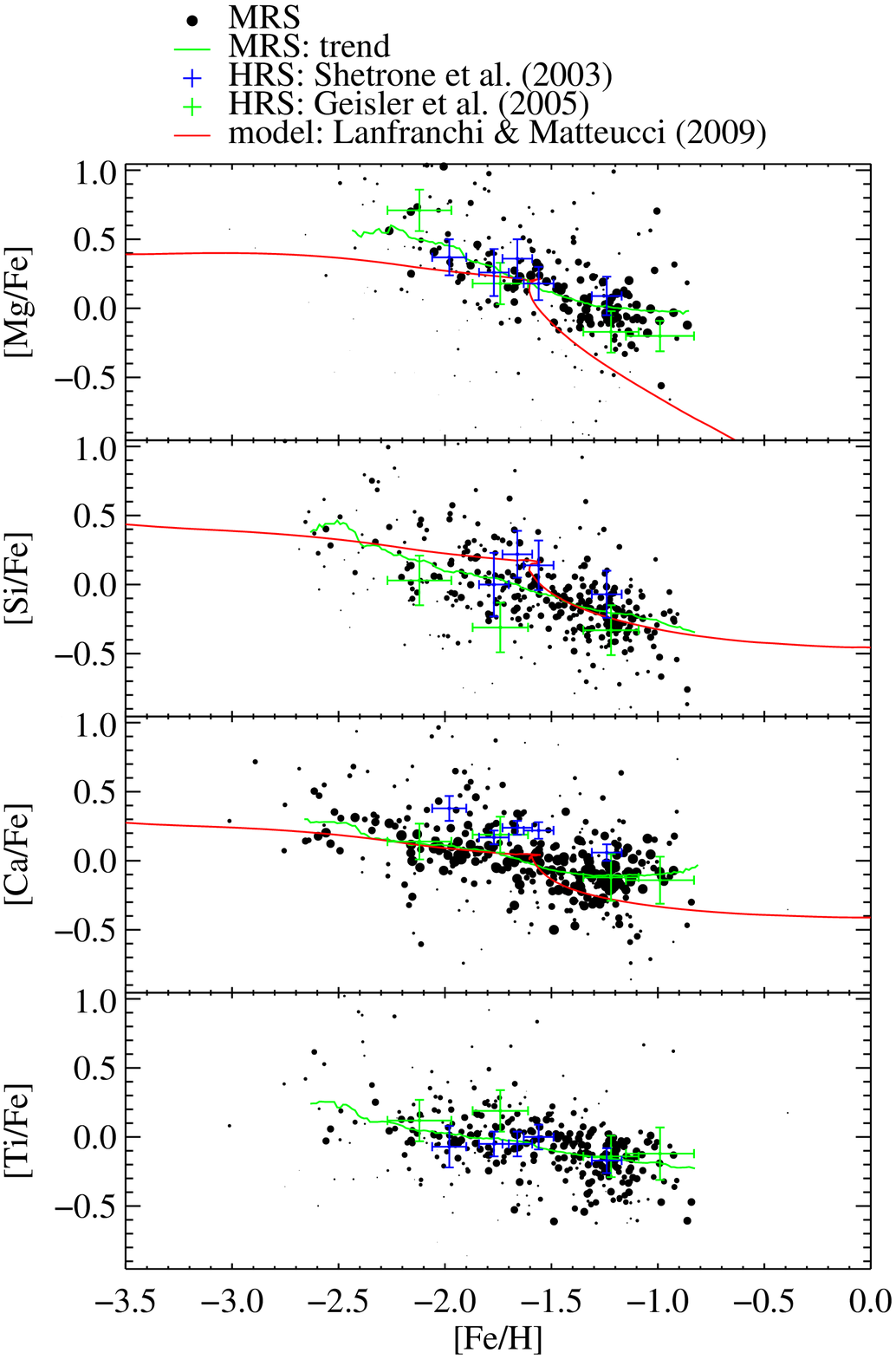}
\caption{As in Fig.~\ref{fig:alphafe_feh_gc}, the black points show
  medium-resolution multi-element abundances in Sculptor.  The points
  with error bars show published high-resolution data
  (\citeauthor{she03} \citeyear{she03}, {\it blue}, and
  \citeauthor{gei05} \citeyear{gei05}, {\it green}).  The green line
  is the inverse variance-weighted average of at least 20 stars within
  a window of $\Delta\mathfeh = 0.25$.  The red line shows the
  chemical evolution model of \citet[][updated 2009]{lan04}.  The
  onset of Type~Ia SNe causes the decline in [X/Fe] with \feh.  Mg
  declines steadily because it is produced exclusively in Type~II SNe,
  but Si, Ca, and Ti are produced in both Type~Ia and II
  SNe.\label{fig:alphafe_feh_lf}}
\end{figure}

\begin{figure}[t!]
\includegraphics[width=\linewidth]{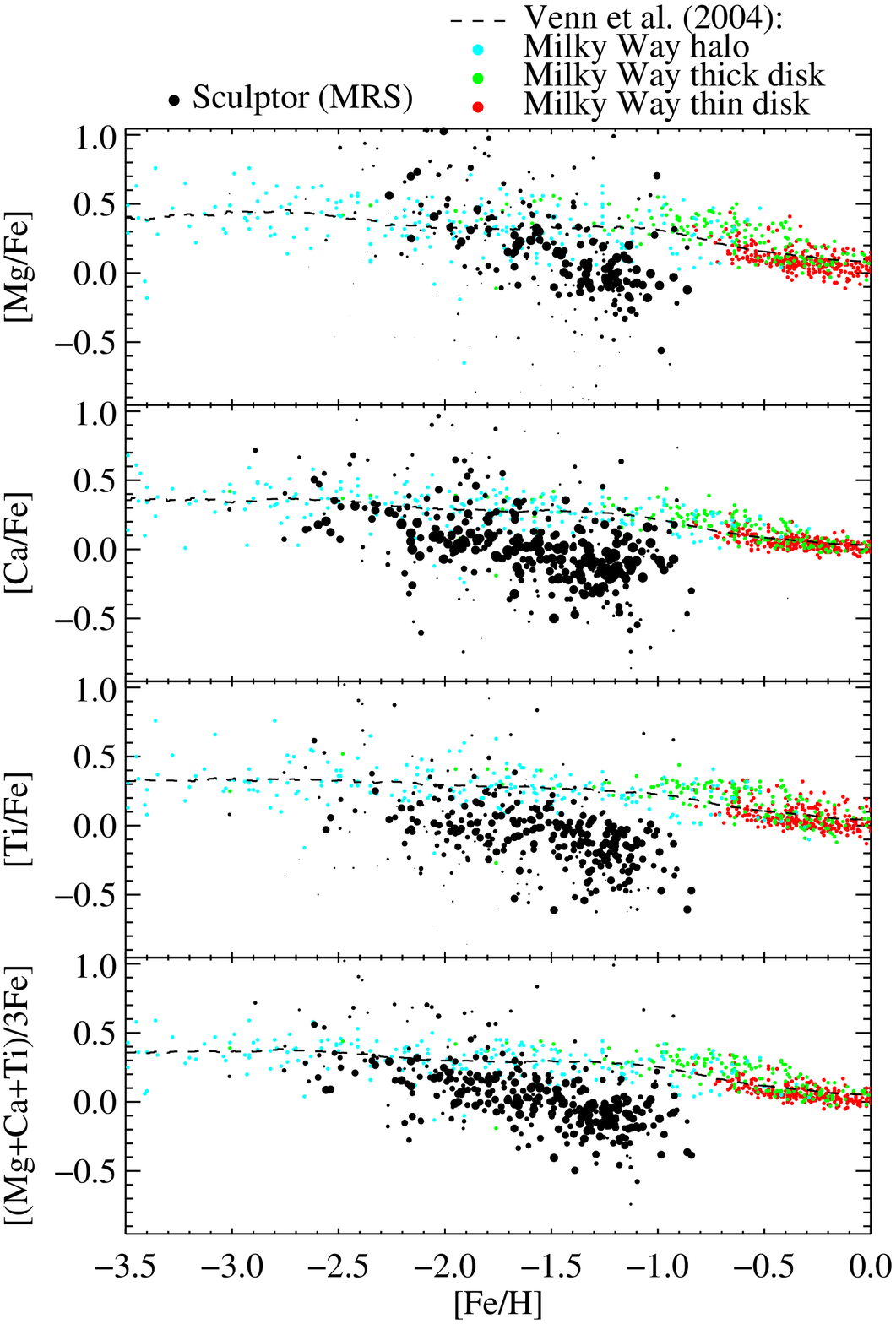}
\caption{As in Fig.~\ref{fig:alphafe_feh_gc}, the black points show
  medium-resolution multi-element abundances in Sculptor.  The colored
  points show different components of the Milky Way \citep{ven04}: the
  thin disk ({\it red}), the thick disk ({\it green}), and the field
  halo ({\it cyan}).  The dashed lines are moving averages of the MW
  data in 0.75~dex bins of \feh.  In Sculptor, \afe\ falls at lower
  \feh\ than in the halo, indicating that the halo field stars were
  less polluted by Type~Ia SNe and therefore formed more rapidly than
  Sculptor stars.\label{fig:alphafe_feh_venn}}
\end{figure}

The discrepancy between halo and dSph abundances extends beyond the
MDF.  In the first HRS study of stars in a dSph, \citet*{she98} found
that the [Ca/Fe] ratio of metal-poor stars in Draco appeared solar, in
contrast to the enhanced halo field stars.  \citet*{she01} and
\citet{she03} confirmed the same result in \object[NAME Sextans
dSph]{Sextans}, Ursa Minor, Sculptor, Fornax, Carina, and \object[NAME
LEO I dSph]{Leo~I}, and they included other $\alpha$ elements in
addition to Ca.


Here, we present the largest sample of \afe\ measurements in any dSph.
Figure~\ref{fig:alphafe_feh_gc} shows [Mg/Fe], [Ca/Fe], and [Ti/Fe]
versus [Fe/H] for Sculptor.  The figure also shows the mean and
standard deviations of all of the individual stellar abundance
measurements for each of the seven GCs in the sample of KGS08.  All of
the modifications to the KGS08 technique described in
Secs.~\ref{sec:prep} and \ref{sec:measure} apply to the GC
measurements in Fig.~\ref{fig:alphafe_feh_gc}.  Although our
discussion in the appendix demonstrates that our measurements are
accurate on an absolute scale by comparing to several different HRS
studies in Sculptor, it is also instructive to compare abundances
measured with the same technique in two types of stellar systems.  All
four element ratios slope downward with \feh\ in Sculptor but remain
flat in the GCs.  Additionally, the larger spread of [Mg/Fe] than
other element ratios in the GCs is not due to larger measurement
uncertainties but to the known intrinsic spread of Mg abundance in
some GCs \citep[see the review by][]{gra04}.  [Si/Fe], [Ca/Fe], and
[Ti/Fe] are more slightly sloped than [Mg/Fe] in Sculptor because both
Type~Ia and Type~II SNe produce Si, Ca, and Ti, but Type~II SNe are
almost solely responsible for producing Mg \citep{woo95}.  Finally, to
maximize the SNR of the element ratio measurements, we average the
four ratios together into one number called \afe.  The \afe\ ratio is
flat across the GCs, but it decreases with increasing \feh\ in
Sculptor.

Quantitative models of chemical evolution in dwarf galaxies are
consistent with these trends.  At a certain time corresponding to a
certain \feh\ in the evolution of the dSph, Type~Ia begin to pollute
the interstellar medium with gas at subsolar \afe.  More metal-rich
stars that form from this gas will have lower \afe\ than the more
metal-poor stars.  \citet{lan04} have developed a sophisticated model
that includes SN feedback and winds.  They predicted the abundance
distributions of six dSphs, including Sculptor.
Figure~\ref{fig:alphafe_feh_lf} shows our measurements with their
predictions (updated with new SN yields, G.~Lanfranchi 2009, private
communication).  As predicted, the range of [Mg/Fe] is larger than the
range of [Ca/Fe] or [Si/Fe] because Mg is produced exclusively in
Type~II SNe whereas Si and Ca are produced in both Type~Ia and II SNe
\citep{woo95}.  We do not observe strong evidence for a predicted
sharp steepening in slope of both elements at $\mathfeh \sim -1.8$,
but observational errors and intrinsic scatter may obscure this
``knee.''  Also, the observed \feh\ at which [Mg/Fe] begins to drop is
higher than the model predicts, indicating a less intense wind than
used in the model.  Note that the element ratios [X/Fe] become
negative (subsolar) at high enough \feh, as predicted by the models.
\citet{lan04} do not predict [Ti/Fe] because it behaves more like an
Fe-peak element than an $\alpha$ element.

In addition to trends of \afe\ with \feh, \citet{mar06,mar08}
predicted the distribution functions of \feh\ and \afe\ of a
Draco-like dSph.  The range of \feh\ they predicted is nearly
identical to the range we observe in Sculptor, and the shapes of both
distributions are similar.  The outcome of the models depends on the
mass of the dSph.  Sculptor is ten times more luminous than Draco
\citep{mat98} and therefore may have a larger total mass.  [However,
\citet{str08} find that all dSphs have the same dynamical mass within
300~pc of their centers.  It is unclear whether the total masses of
the original, unstripped dark matter halos are the same.]  In
principle, these chemical evolution models could be used to measure
the time elapsed since different epochs of star formation and their
durations.  We defer such an analysis until the advent of a model
based on a Sculptor-like luminosity or mass.

In Fig.~\ref{fig:alphafe_feh_venn}, we compare individual stellar
abundances in Sculptor to MW halo and disk field stars
\citep[compilation by][]{ven04}.  As has been seen in many previous
studies of individual stellar abundances in dSphs, \afe\ falls at a
significantly lower \feh\ in Sculptor than in the MW halo.  The drop
is particularly apparent in [Mg/Fe], which is the element ratio most
sensitive to the ratio of the contributions of Type~II to Type~Ia SNe.
The other element ratios also drop sooner in Sculptor than in the
halo, but appear lower than in the halo at all metallicities.  Along
with the MDF comparison in Sec.~\ref{sec:halomdf}, this result is
consistent with the suggestion by \citet{rob05} that galaxies
significantly more massive than Sculptor built the inner MW halo.
Their greater masses allowed them to retain more gas and experience
more vigorous star formation.  By the time Type~Ia SNe diluted \afe\
in the massive halo progenitors, the metallicity of the star-forming
gas was already as high as $\mathfeh = -0.5$.  In Sculptor, the
interstellar \feh\ reached only $-1.5$ before the onset of Type~Ia SNe
pollution.

%
%
%

\subsection{Radial Abundance Distributions}
\label{sec:gradients}

\begin{figure}[t!]
\includegraphics[width=\linewidth]{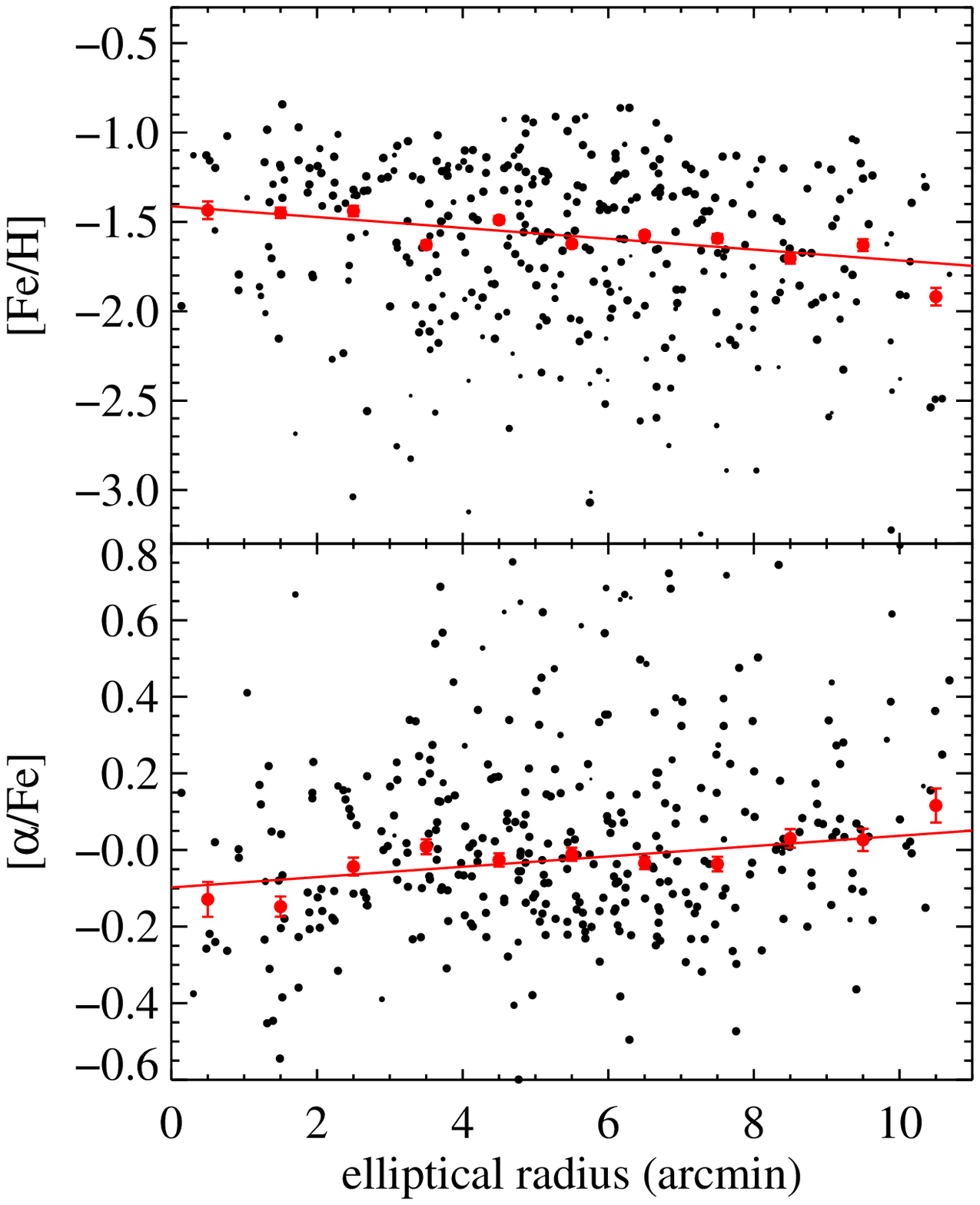}
\caption{Spatial abundance distributions in Sculptor.  Point sizes are
  larger for stars with smaller measurement uncertainties.  The red
  points reflect the mean values in 1~arcmin bins, along with the
  errors on the means.  The red lines are the least-squares linear
  fits.  We detect a gradient of $\sclfehslope \pm \sclfehslopeerr$~dex per
  arcmin in \feh\ and $\sclafeslope \pm \sclafeslopeerr$~dex per arcmin in
  \afe.\label{fig:feh_dist}}
\end{figure}

Because dSphs interact with the MW, they can lose gas through tidal or
ram pressure stripping \citep{lin83}.  The gas preferentially leaves
from the dSph's outskirts, where the gravitational potential is
shallow.  If the dSph experiences subsequent star formation, it must
occur in the inner regions where gas remains.  Sculptor's MDF suggests
a history of extended star formation.  Sculptor might then be expected
to exhibit a radial abundance gradient in the sense that the inner
parts of the dSph are more metal-rich than the outer parts.

The detection of a radial metallicity gradient in Sculptor has been
elusive.  In a photometric study, \citet{hur00} found no evidence for
an age or metallicity gradient.  Based on HRS observations of five
stars \citep[the same sample as][]{she03}, \citet{tol03} found no
correlation between \feh\ and spatial position.  Finally, in a sample
of 308 stars with CaT-based metallicities, T04 detected a significant
segregation in Sculptor: a centrally concentrated, relatively
metal-rich component and an extended, relatively metal-poor component.
\citet{wes06} arrived at the same conclusion, and \citet{wal09}
confirmed the existence of a \feh\ gradient in a sample of 1365
Sculptor members.

In order to detect a gradient, those studies targeted Sculptor stars
at distances of more than 20~arcmin.  The maximum elliptical radius of
this study is 11~arcmin.  Therefore, this study is not ideally
designed to detect radial gradients.  Figure~\ref{fig:feh_dist} shows
the radial distribution of \feh\ and \afe\ in Sculptor.  The $x$-axis
is the length of the semi-major axis of an ellipse defined by
Sculptor's position angle and ellipticity \citep{mat98}.  Although
this study is limited in the spatial extent of targets, we do detect a
gradient of $\sclfehslope \pm \sclfehslopeerr$~dex per arcmin.  This
estimate is very close to the gradient observed by T04.  \citet{wal09}
measure a shallower gradient, but they present their results against
circular radius instead of elliptical radius.

\citet{mar08} predict radial gradients in both \feh\ and \afe\ in
dSphs.  In particular, they expect shallower \feh\ gradients for
longer durations of star formation.  The gradient we observe is
stronger than any of their models.  They also expect very few stars
with low \afe\ at large radius.  Given that \afe\ decreases with \feh\
and \feh\ decreases with distance, it seems reasonable to expect that
\afe\ increases with radius.  In fact, we detect an \afe\ gradient of
$\sclafeslope \pm \sclafeslopeerr$~dex per arcmin.


\section{Conclusions}
\label{sec:concl}

Sculptor is one of the best-studied dwarf spheroidal satellites of the
Milky Way.  In the past ten years, at least five spectroscopic
campaigns at both low and high resolution have targeted this galaxy.
More than any other dSph, Sculptor has aided in the understanding of
the chemical evolution of dSphs and the construction of the Milky Way
stellar halo.

We have sought to increase the sample of multi-element abundances in
Sculptor through MRS.  The advantages over HRS include higher
throughput per resolution element, the ability to target fainter
stars, and multiplexing.  The large sample sizes will enable detailed
comparisons to chemical evolution models of \afe\ and \feh\ in dSphs.
The disadvantages include larger uncertainties, particularly for
elements with few absorption lines in the red, and the inability to
measure many elements accessible to HRS.  MRS is not likely to soon
provide insight into the evolution of neutron-capture elements in
dSphs.

In order to make the most accurate measurements possible, we have made
a number of improvements to the technique of \citet*{kir08a}.  We have
consulted independent HRS of the same stars to confirm the accuracy of
our measurements of \feh, [Mg/Fe], [Ca/Fe], and [Ti/Fe].  In the case
of \feh\ and the average \afe\ our MRS measurements are only slightly
more uncertain than HRS measurements.

Some of the products of this study include

\begin{enumerate}
\item {\bf An unbiased metallicity distribution for Sculptor.}
  Because the synthesis-based abundances do not rely on any empirical
  calibration, their applicability is unrestricted with regard to
  \feh\ range.  The MDF is asymmetric with a long, metal-poor tail, as
  predicted by chemical evolution models of dSphs.  Furthermore, fits
  to simple chemical evolution models shows that Sculptor's MDF is
  consistent with a model that requires no pre-enrichment.

\item {\bf The largest sample of \afe\ and \feh\ measurements in any
  single dSph: \sclnmember\ stars.} We have confirmed the trend for \afe\
  to decrease with \feh, as shown by \citet{gei07} with just nine
  stars from the studies of \citet{she03} and \citet{gei05}.  Chemical
  evolution models may be constructed from these measurements to
  quantify the star formation history of Sculptor.

\item {\bf The detection of radial \feh\ and \afe\ gradients.}  Our
  sample probes a smaller range than previous studies; nonetheless, we
  find a $\sclfehslope \pm \sclfehslopeerr$~dex per arcmin gradient in \feh\
  and a $\sclafeslope \pm \sclafeslopeerr$~dex per arcmin gradient in \afe.

\item {\bf The discovery of a Sculptor member star with
  $\rm{\bf{[Fe/H]}} \mathbf{= \sclempfehtwo \pm \sclempfeherrtwo}$.}
  This discovery suggests that since-disrupted galaxies similar to
  Sculptor may have played a role in the formation of the Milky Way
  metal-poor halo.  High-resolution spectroscopy of individual stars
  will confirm or refute this indication.
\end{enumerate}

Much more can be done with this technique in other galaxies.  The
stellar population of a dSph depends heavily on its stellar mass.  For
instance, \citet{lan04} and \citet{rob05} predict that more massive
satellites have an \afe\ ``knee'' at higher \feh.  In the next papers
in this series, we intend to explore the multi-element abundance
distributions of other dSphs and compare them to each other.  We will
observe how the shapes of the MDFs and the \afe--\feh\ diagrams change
with dSph luminosity or stellar mass.  These observations should aid
our understanding of star formation, chemical evolution, and the
construction of the Galaxy.

\acknowledgments We thank Kyle Westfall for providing the photometric
catalog, Gustavo Lanfranchi and Francesca Matteucci for providing
their chemical evolution model, David Lai for thoughtful
conversations, and the anonymous referee for helpful comments that
improved this manuscript.  The generation of synthetic spectra made
use of the Yale High Performance Computing cluster Bulldog.  ENK is
grateful for the support of a UC Santa Cruz Chancellor's Dissertation
Year Fellowship.  PG acknowledges NSF grant AST-0307966, AST-0607852,
and AST-0507483.  CS acknowledges NSF grant AST-0607708.

{\it Facility:} \facility{Keck:II (DEIMOS)}


\appendix
\section{Accuracy of the Abundance Measurements}
\label{sec:accuracy}

In order to quantify the accuracy of the MRS measurements, we examine
the spectra of stars observed more than once and stars with previous
HRS measurements.

\subsection{Duplicate Observations}
\label{sec:duplicate}

\begin{figure}[t!]
\centering
\begin{minipage}[t]{0.49\textwidth}
\centering
\includegraphics[width=\linewidth]{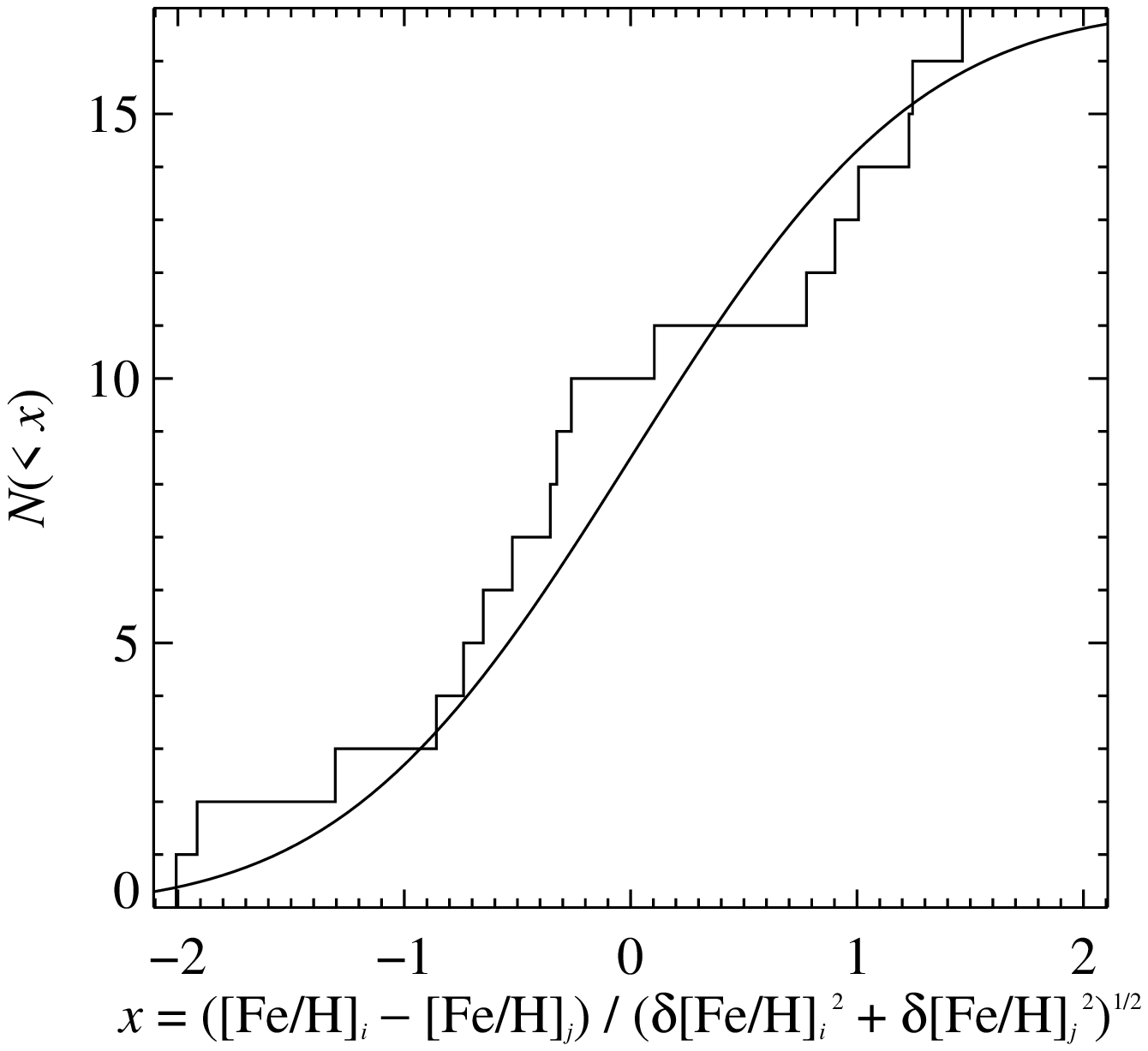}
\caption{Cumulative distribution of differences between the repeat
  measurements of \feh\ for \sclndup\ stars divided by the estimated
  error of the difference.  The curve is the integral of a unit
  Gaussian.  The curve matches the distribution well, indicating that
  the errors are estimated properly.\label{fig:fehdup}}
\end{minipage}
\begin{minipage}[t]{0.49\textwidth}
\centering
\includegraphics[width=\linewidth]{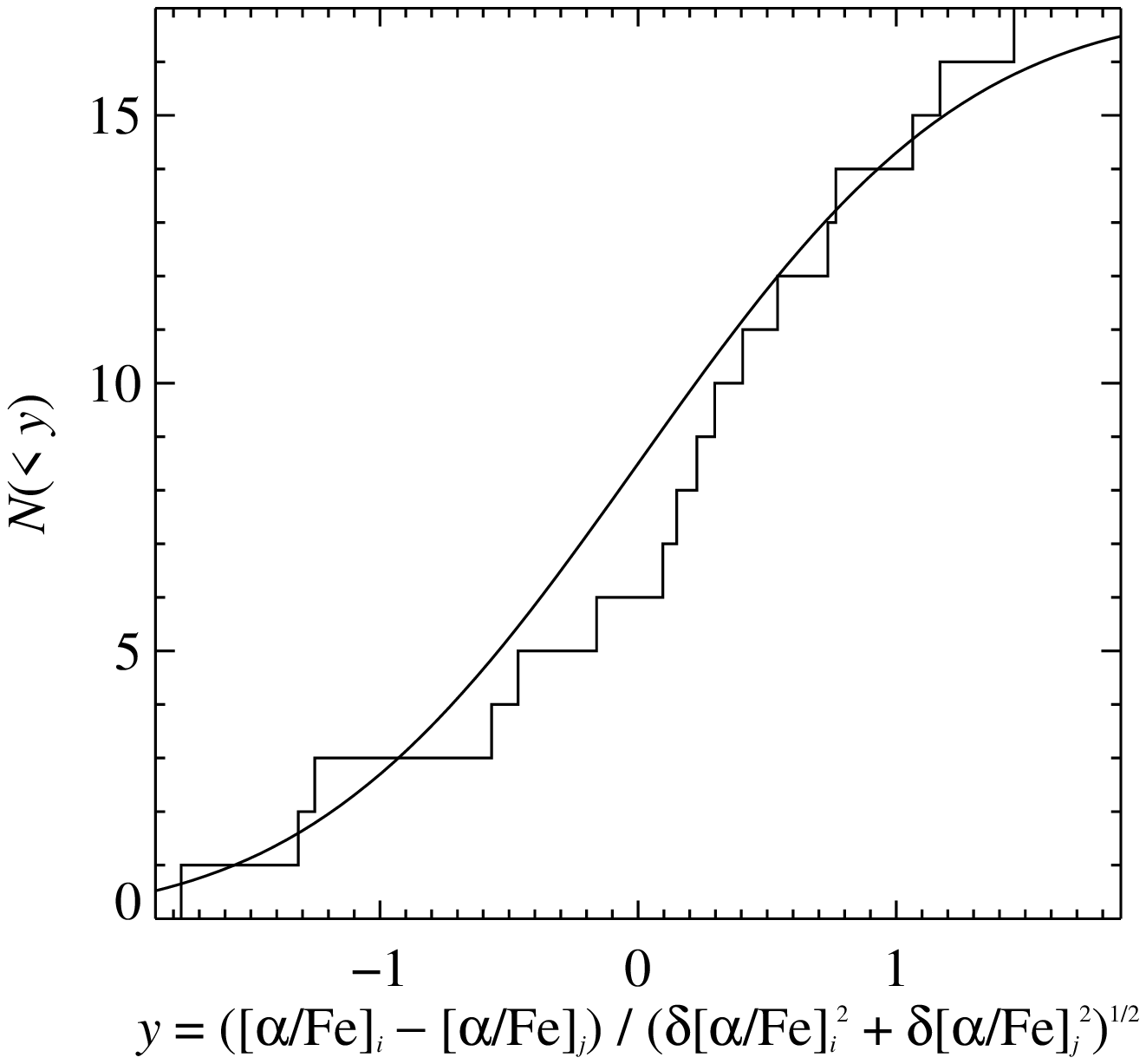}
\caption{Same as Fig.~\ref{fig:fehdup} for \afe, which is the average
  of [Mg/Fe], [Si/Fe], [Ca/Fe], and [Ti/Fe].\label{fig:alphadup}}
\end{minipage}
\end{figure}

The repeat observations of \sclndup\ stars provide insight on the effect
of random error on the measurements of \feh\ and \afe.
Figures~\ref{fig:fehdup} and \ref{fig:alphadup} summarize the
comparisons of measurements of different spectra of the same stars.
They show the cumulative distribution of the absolute difference
between the measured \feh\ and \afe\ for each pair of spectra divided
by the expected error of the difference (see Sec.~\ref{sec:error}).
The solid curve is the integral of a unit Gaussian, which represents
the expected cumulative distribution if the estimated errors
accurately represent the true measurement errors.  In calculating the
expected error of the difference, we apply the systematic error to
only one of the two stars.  Even though the same technique is used to
measure abundances in both stars, some systematic error is appropriate
because the wavelength range within a pair of spectra differs by
300--400~\AA.  The different Fe lines in these ranges span a different
range of excitation potentials, and the Levenberg-Marquardt algorithm
converges on different solutions.


\subsection{Comparison to High-Resolution Measurements}
\label{sec:hrscompare}

\begin{deluxetable*}{lcccccccc}
\tablewidth{0pt}
\tablecolumns{9}
\tablecaption{Abundances of Stars with Previous High-Resolution Spectroscopy\label{tab:hrsabund}}
\tablehead{\colhead{star} & \colhead{$T_{\rm{eff}}$} & \colhead{$\log g$} & \colhead{$\xi$} & \colhead{[Fe/H]} & \colhead{[Mg/Fe]} & \colhead{[Si/Fe]} & \colhead{[Ca/Fe]} & \colhead{[Ti/Fe]} \\
\colhead{ } & \colhead{(K)} & \colhead{(cm~s$^{-2}$)} & \colhead{(km~s$^{-1}$)} & \colhead{(dex)} & \colhead{(dex)} & \colhead{(dex)} & \colhead{(dex)} & \colhead{(dex)}}
\startdata
\cutinhead{Previous High-Resolution Measurements}
H482 & 4400 & 1.10 & 1.70 & $-1.24 \pm 0.07$ & $+0.09 \pm 0.14$ & $-0.07 \pm 0.17$ & $+0.06 \pm 0.06$ & $-0.17 \pm 0.09$ \\
H459 & 4500 & 1.00 & 1.65 & $-1.66 \pm 0.07$ & $+0.36 \pm 0.14$ & $+0.22 \pm 0.17$ & $+0.24 \pm 0.05$ & $-0.05 \pm 0.09$ \\
H479 & 4325 & 0.70 & 1.70 & $-1.77 \pm 0.07$ & $+0.26 \pm 0.17$ & $+0.00 \pm 0.23$ & $+0.17 \pm 0.05$ & $-0.05 \pm 0.09$ \\
H400 & 4650 & 0.90 & 1.70 & $-1.98 \pm 0.08$ & $+0.37 \pm 0.13$ &     \nodata      & $+0.38 \pm 0.09$ & $-0.07 \pm 0.15$ \\
H461 & 4500 & 1.20 & 1.70 & $-1.56 \pm 0.07$ & $+0.18 \pm 0.12$ & $+0.14 \pm 0.18$ & $+0.22 \pm 0.06$ & $+0.00 \pm 0.09$ \\
1446 & 3900 & 0.00 & 2.30 & $-1.22 \pm 0.13$ & $-0.17 \pm 0.15$ & $-0.33 \pm 0.18$ & $-0.12 \pm 0.17$ & $-0.14 \pm 0.15$ \\
195  & 4250 & 0.20 & 1.80 & $-2.12 \pm 0.15$ & $+0.71 \pm 0.15$ & $+0.03 \pm 0.18$ & $+0.14 \pm 0.13$ & $+0.12 \pm 0.15$ \\
982  & 4025 & 0.50 & 2.20 & $-0.99 \pm 0.16$ & $-0.20 \pm 0.11$ &     \nodata      & $-0.14 \pm 0.17$ & $-0.12 \pm 0.19$ \\
770  & 4075 & 0.00 & 1.90 & $-1.74 \pm 0.13$ & $+0.18 \pm 0.15$ & $-0.31 \pm 0.18$ & $+0.19 \pm 0.13$ & $+0.19 \pm 0.15$ \\
\cutinhead{Medium-Resolution Measurements}
H482 & 4347 & 0.83 & 1.95 & $-1.24 \pm 0.14$ & $-0.03 \pm 0.17$ & $-0.16 \pm 0.20$ & $-0.19 \pm 0.13$ & $-0.12 \pm 0.11$ \\
H459 & 4390 & 1.12 & 1.88 & $-1.88 \pm 0.14$ &     \nodata      & $-0.41 \pm 0.48$ & $+0.34 \pm 0.27$ & $+0.08 \pm 0.18$ \\
H479 & 4271 & 0.63 & 1.99 & $-1.79 \pm 0.14$ & $+0.27 \pm 0.42$ & $-0.22 \pm 0.22$ & $+0.24 \pm 0.18$ & $-0.13 \pm 0.11$ \\
H400 & 4692 & 1.36 & 1.82 & $-1.97 \pm 0.15$ &     \nodata      & $+0.58 \pm 0.23$ & $+0.90 \pm 0.65$ & $+0.10 \pm 0.19$ \\
H461 & 4313 & 0.78 & 1.96 & $-1.81 \pm 0.15$ & $+0.36 \pm 0.61$ & $+0.05 \pm 0.32$ & $+0.39 \pm 0.32$ & $+0.25 \pm 0.16$ \\
1446 & 3838 & 0.49 & 2.03 & $-1.22 \pm 0.14$ & $-0.03 \pm 0.20$ & $-0.07 \pm 0.21$ & $-0.05 \pm 0.24$ & $-0.40 \pm 0.11$ \\
195  & 4308 & 0.65 & 1.99 & $-2.05 \pm 0.14$ & $+0.41 \pm 0.17$ & $+0.06 \pm 0.19$ & $+0.09 \pm 0.10$ & $+0.02 \pm 0.11$ \\
982  & 4147 & 0.52 & 2.02 & $-0.84 \pm 0.14$ &     \nodata      &     \nodata      & $-0.30 \pm 0.23$ & $-0.47 \pm 0.11$ \\
770  & 4247 & 0.59 & 2.00 & $-1.62 \pm 0.14$ & $+0.13 \pm 0.29$ & $-0.23 \pm 0.21$ & $+0.02 \pm 0.11$ & $-0.04 \pm 0.11$ \\
\enddata
\end{deluxetable*}

The most reliable test of the MRS atmospheric parameter and abundance
estimates is to compare with completely independent observations and
analyses of the same stars.  Table~\ref{tab:hrsabund} lists the
previous HRS measurements of nine Sculptor members \citep{she03,gei05}
as well as the DEIMOS measurements of the same stars.  Unfortunately,
these two HRS studies share no stars in common and therefore cannot be
compared with each other.

\begin{figure}[t!]
\centering
\begin{minipage}[t]{0.49\textwidth}
\centering
\includegraphics[width=\linewidth]{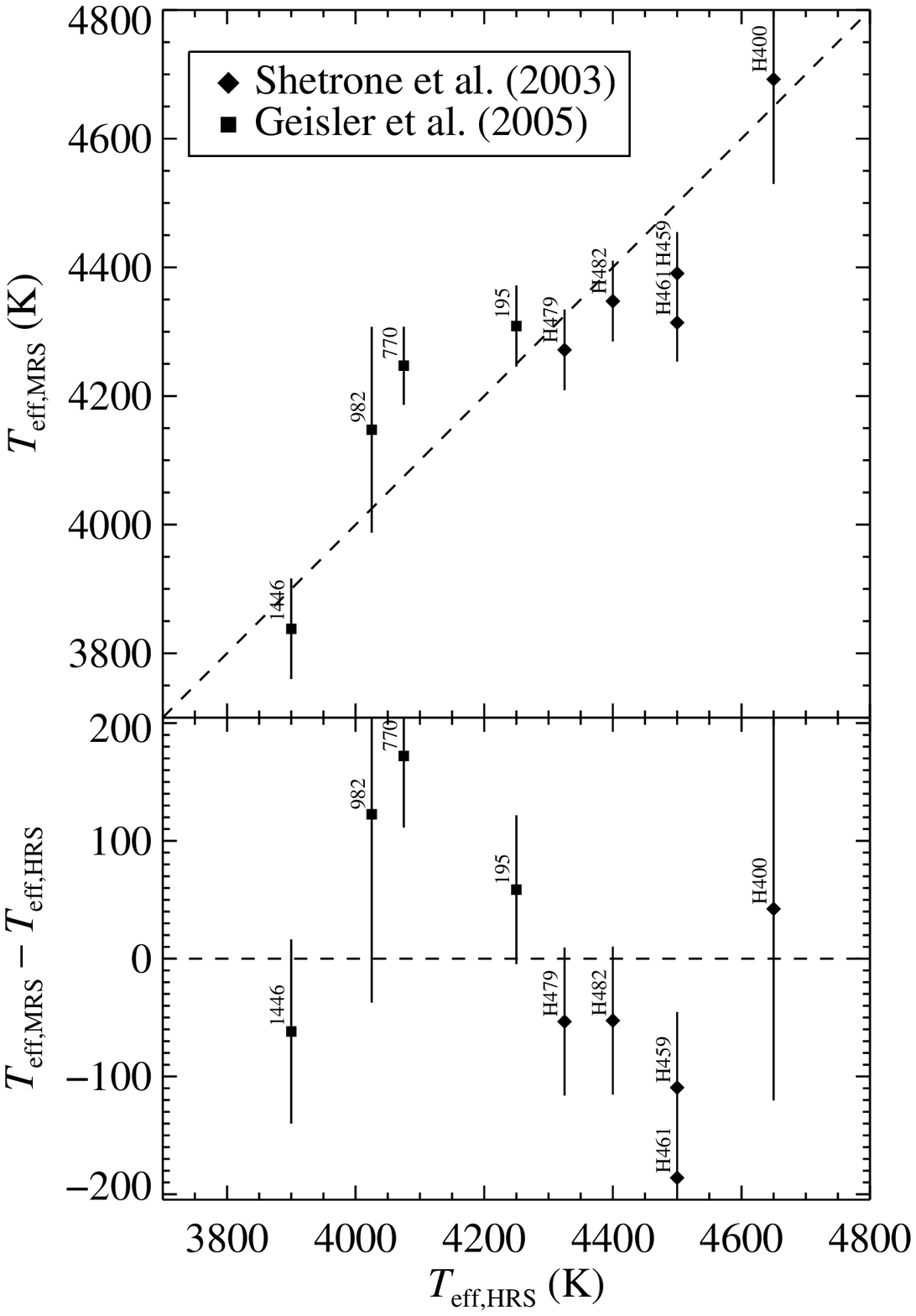}
\caption{Comparison between effective temperature (\teff) used in
  previous HRS abundance analyses and photometric \teff\ used for this
  work's MRS abundance analysis.  Symbol shape indicates the reference
  for the HRS abundances.  Star names from Tab.~\ref{tab:hrslist} are
  printed to the upper left of each point.\label{fig:hrsteff}}
\end{minipage}
\begin{minipage}[t]{0.49\textwidth}
\centering
\includegraphics[width=\linewidth]{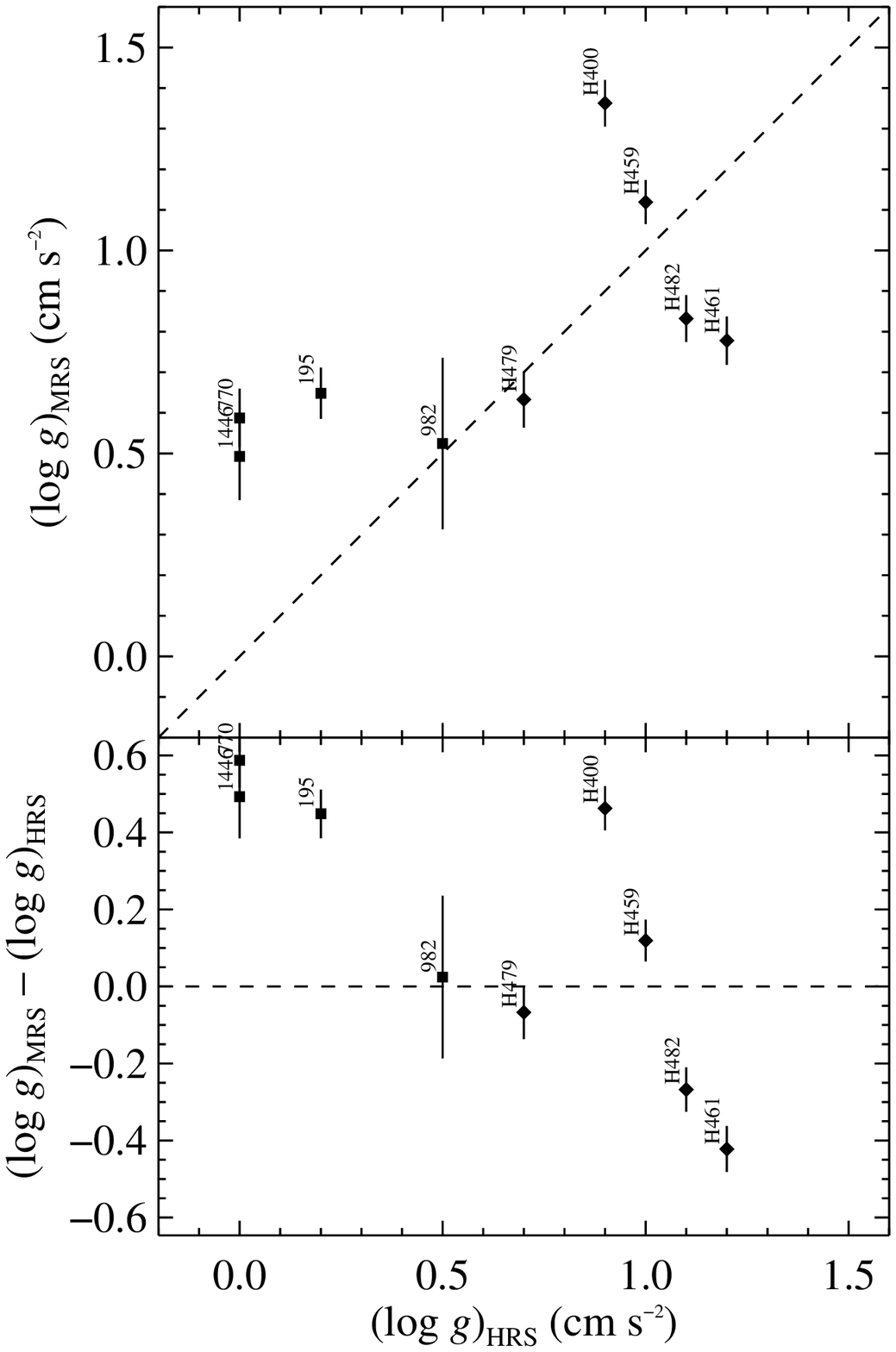}
\caption{Same as Fig.~\ref{fig:hrsteff} except for surface gravity
  (\logg).  The error bars represent photometric error, and they are
  given by replacing \teff\ with \logg\ in
  Eq.~\ref{eq:tefferr}.\label{fig:hrslogg}}
\end{minipage}
\end{figure}

\begin{figure}[t!]
\centering
\begin{minipage}[t]{0.49\textwidth}
\centering
\includegraphics[width=\linewidth]{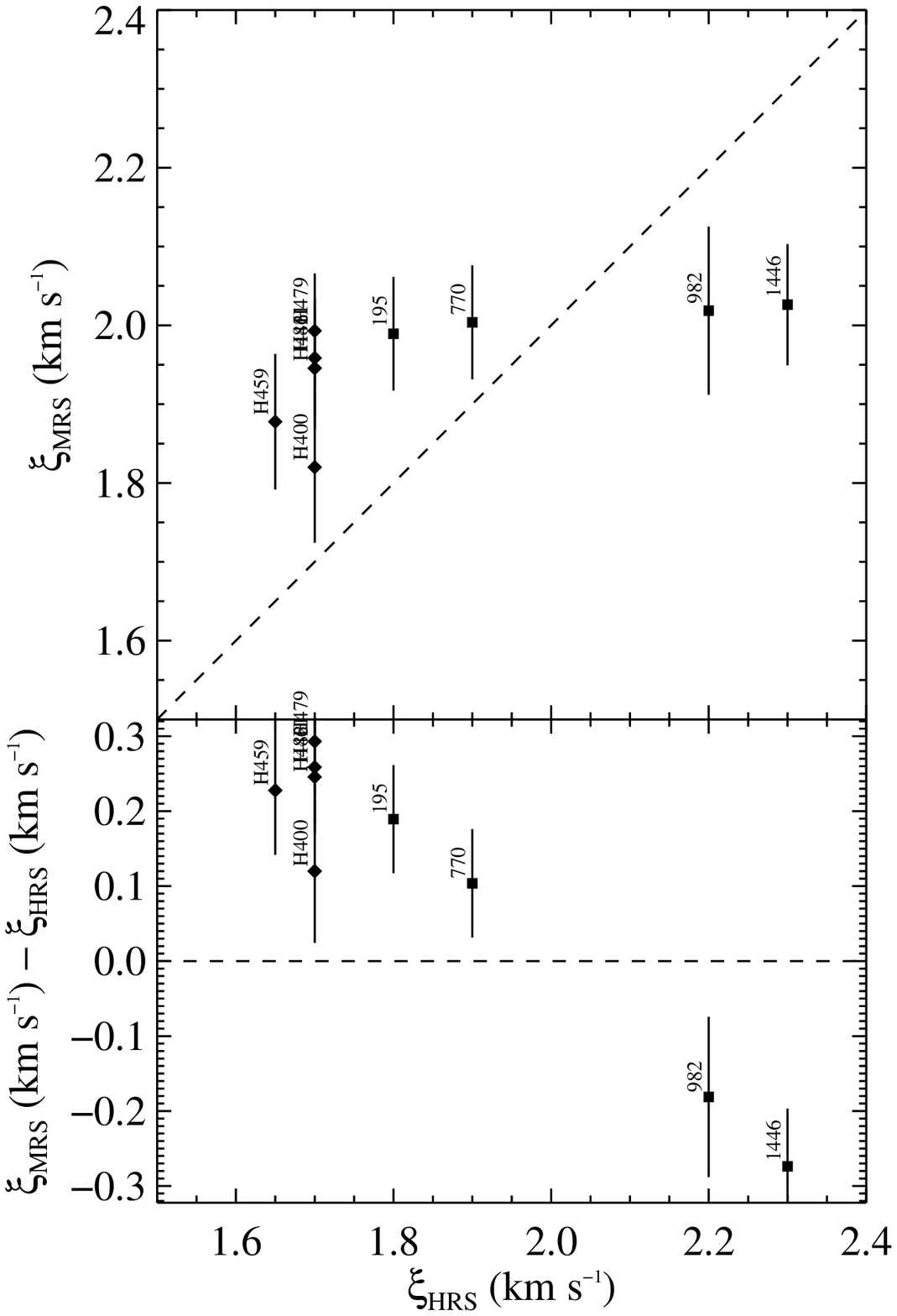}
\caption{Same as Fig.~\ref{fig:hrsteff} except for microturbulent
  velocity ($\xi$).  The error bars are found by propagating the error
  on \logg\ through Eq.~\ref{eq:vtlogg}.\label{fig:hrsvt}}
\end{minipage}
\begin{minipage}[t]{0.49\textwidth}
\centering
\includegraphics[width=\linewidth]{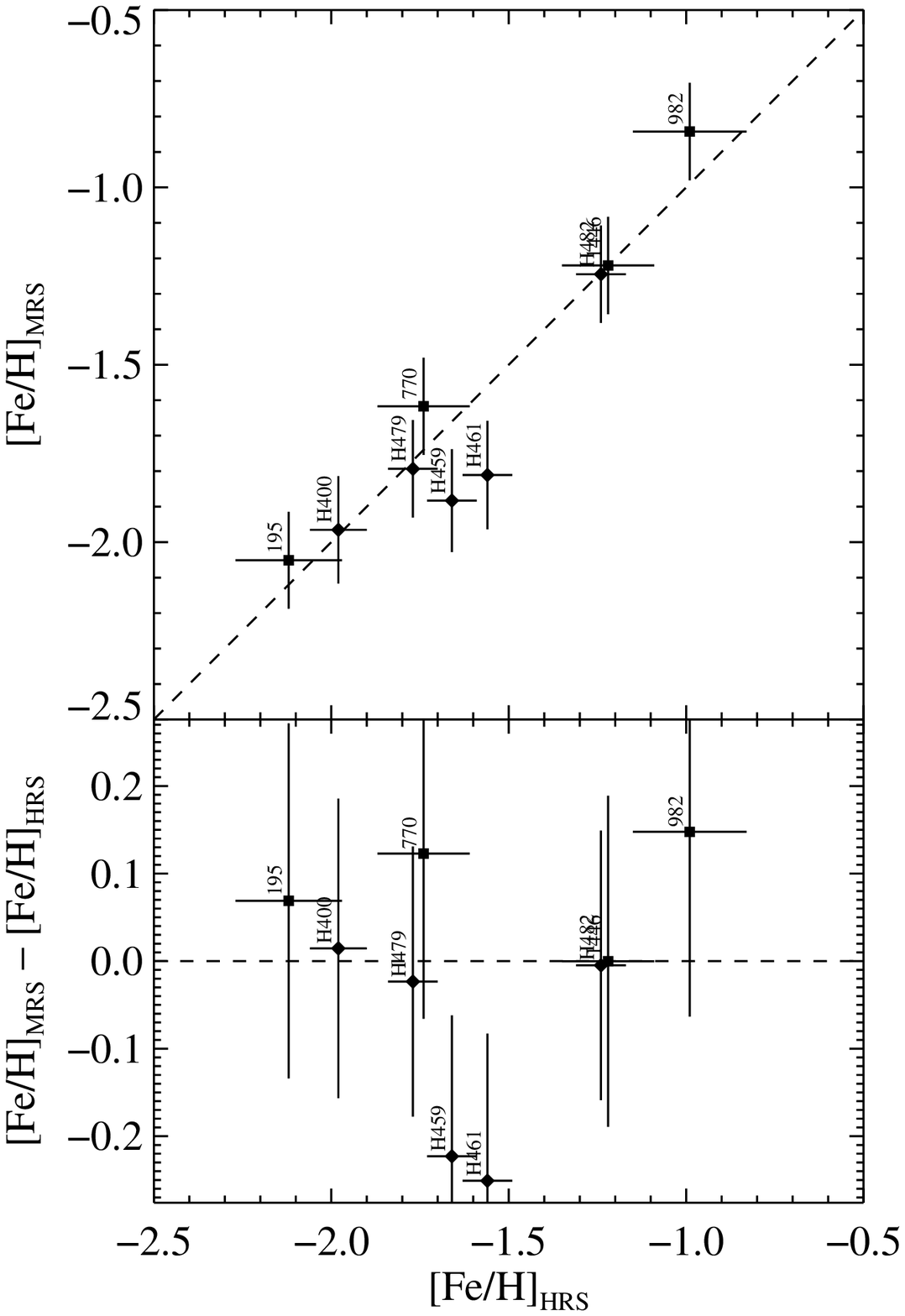}
\caption{Comparison between \feh\ derived from previous HRS abundance
  analyses and \feh\ derived from this work's MRS abundance analysis.
  Symbols are the same as in
  Fig.~\ref{fig:hrsteff}.\label{fig:hrsfeh}}
\end{minipage}
\end{figure}

Of \teff, \logg, and $\xi$, any spectroscopic abundance measurement is
most sensitive to \teff.  In general, underestimating \teff\ leads to
an underestimate of \feh.  \citet[][hereafter S03]{she03} determine
\teff\ spectroscopically by minimizing the slope of the derived
abundance for each line versus excitation potential.
\citet[][hereafter G05]{gei05} determine \teff\ photometrically with
empirical color-temperature relations.  Figure~\ref{fig:hrsteff} shows
\teff\ from those studies and this one for each of the nine stars in
common.  The MRS temperatures do not follow the temperatures of either
HRS study better than the other.

Both S03 and G05 measure \logg\ spectroscopically from demanding
ionization equilibrium: \feh\ measured from \ion{Fe}{1} lines must
match that measured from \ion{Fe}{2} lines.  However, our red spectra
have very few measurable \ion{Fe}{2} lines.  Alternatively, \logg\ may
be determined from a star's absolute magnitude and \teff\ via the
Stefan-Boltzmann law.  Even though gravity depends on the inverse
square of \teff\ and the inverse square root of luminosity, luminosity
imposes a stronger constraint on \logg\ because of its larger range on
the RGB than \teff.  Even accounting for the error in the distance
modulus to Sculptor, the typical error on photometric \logg\ is $\sim
0.1$~dex.  Therefore, we determine \logg\ from photometry alone.
Figure~\ref{fig:hrslogg} shows the comparison between \logg\ used by
S03 and G05 and this study.  The agreement is not particularly good,
with discrepancies up to 0.6~dex.  However, the photometric values of
\logg\ are more accurate than can be determined from the
medium-resolution red spectra, which show very few lines of ionized
species.  Furthermore, as discussed below, errors in \logg\ influence
the abundance measurements much less than errors in \teff.

Both S03 and G05 measure microturbulent velocity ($\xi$) by forcing
all Fe lines to give the same abundance regardless of their reduced
width.  We have fixed $\xi$ to \logg\ with an empirical relation
(Eq.~\ref{eq:vtlogg}).  Figure~\ref{fig:hrsvt} compares the HRS
microturbulent velocities ($\xi$) to our adopted values.  The largest
discrepancy is 0.3~km~s$^{-1}$.

Figure~\ref{fig:hrsfeh} shows the comparison between HRS and MRS
\feh\ measurements for the same stars.  The agreement is very good
($\sigma = \sclfehhrssigma$~dex).  Just two stars out of nine do not fall
within $1\sigma$ of the one-to-one line.

The MRS \feh\ for star 770 is larger than the HRS \feh.  The MRS
\teff\ is also significantly larger than the HRS \teff\ for this star.
Similarly, the MRS \teff\ for star H461 is lower than the HRS \teff,
forcing the MRS \feh\ lower than the HRS \feh.  In fact, even the
smaller deviations from the \feh\ one-to-one line can be attributed to
deviations from the \teff\ one-to-one line.  No such correlation can
be attributed to deviations in \logg\ or $\xi$.  The close
correspondence between Figs.~\ref{fig:hrsteff} and \ref{fig:hrsfeh}
demonstrates that \teff\ is the dominant atmospheric parameter in
determining metallicity.

\begin{figure}[t!]
\centering
\includegraphics[width=0.5\textwidth]{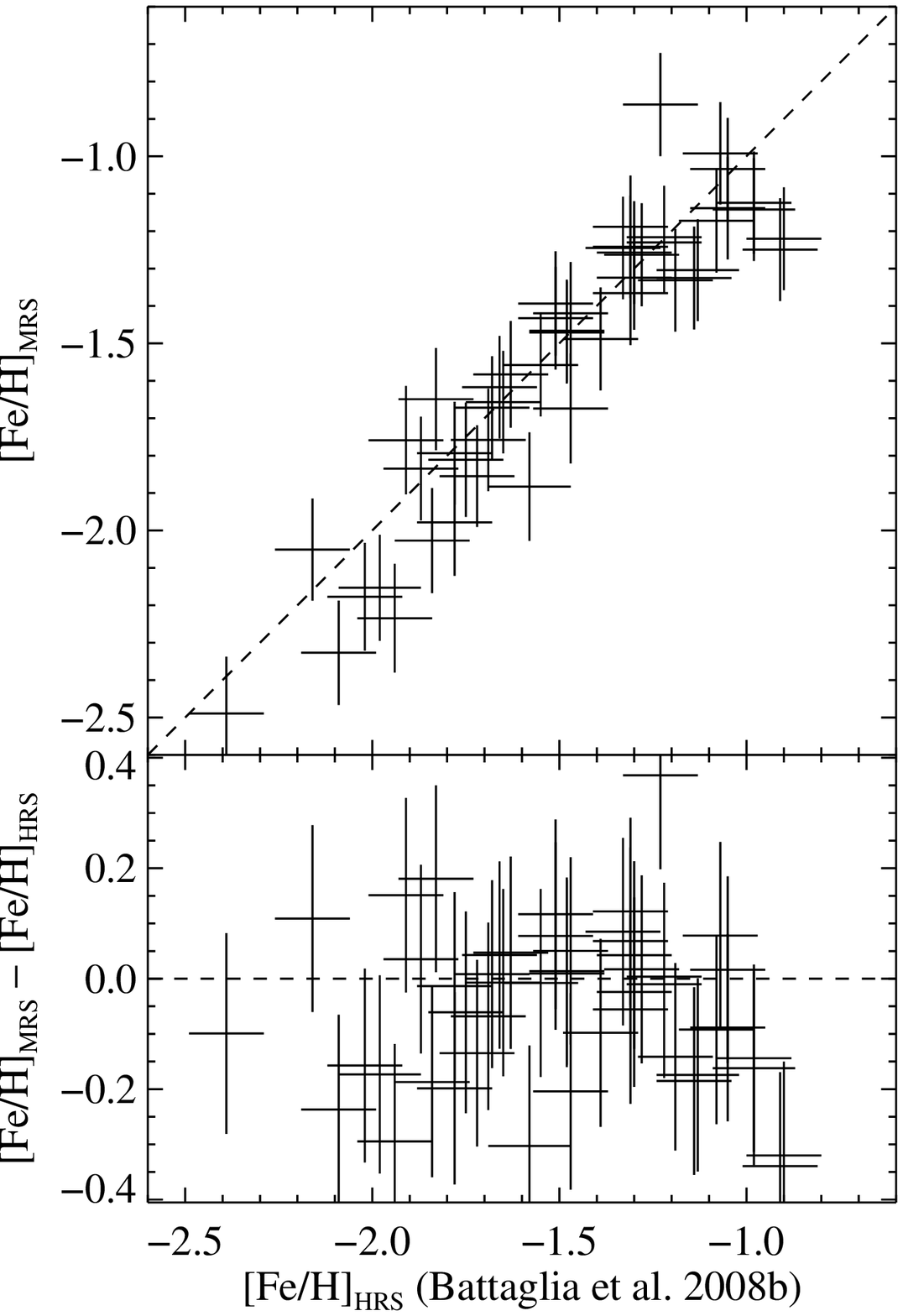}
\caption{Comparison between the HRS spectral synthesis measurements of
  \feh\ of \citet{bat08b} and the synthesis-based medium resolution
  measurements of \feh\ (this work) for the stars observed in both
  studies.\label{fig:bat08feh}}
\end{figure}

B08b published a catalog of VLT/FLAMES \feh\ measurements based on
both the EW of the infrared \ion{Ca}{2} triplet (CaT) and HRS (Hill et
al., in preparation).  The two resolution modes of FLAMES ($R \sim
6500$ and $R \sim 20 000$) allowed them to complete both MRS and HRS
analyses with the same instrument.  Their high-resolution
spectroscopic sample and ours overlap by \sclnbat\ stars, which are
shown in Fig.~\ref{fig:bat08feh}.  The agreement ($\sigma =
\sclfehbatsigma$~dex) is as good as the previous comparison to HRS
studies.

The B08b HRS measurements rely on atmospheric parameters determined
from both five-band photometry and spectroscopy.  We also measure
\teff\ spectrophotometrically.  Our methods may be similar, although
we do not use infrared photometry.  There appears to be a small
systematic trend such that our MRS measurements are lower than the
B08b HRS measurements of \feh\ at both low and high \feh.  The average
discrepancy at the extrema of the residuals is 0.2~dex.  We withhold a
detailed investigation of these residuals until publication of the
details of the HRS study.

B08b share seven stars in common with S03 and G05.  To emphasize the
accuracy of our MRS analysis, we note that the scatter of the
differences between the two sets of HRS studies ($\sigma =
\sclfehbathrssigma$~dex) is in fact larger than the scatter in the
comparison between the MRS \feh\ and the same seven stars of S03 and
G05 ($\sigma = \sclfehcuthrssigma$~dex).  This small sample does not
indicate that the MRS measurements are more accurate than any HRS
measurements, but it does suggest that the accuracy is competitive.

\begin{figure}[t!]
\centering
 \leavevmode
 \columnwidth=.45\columnwidth
 \includegraphics[width=0.45\textwidth]{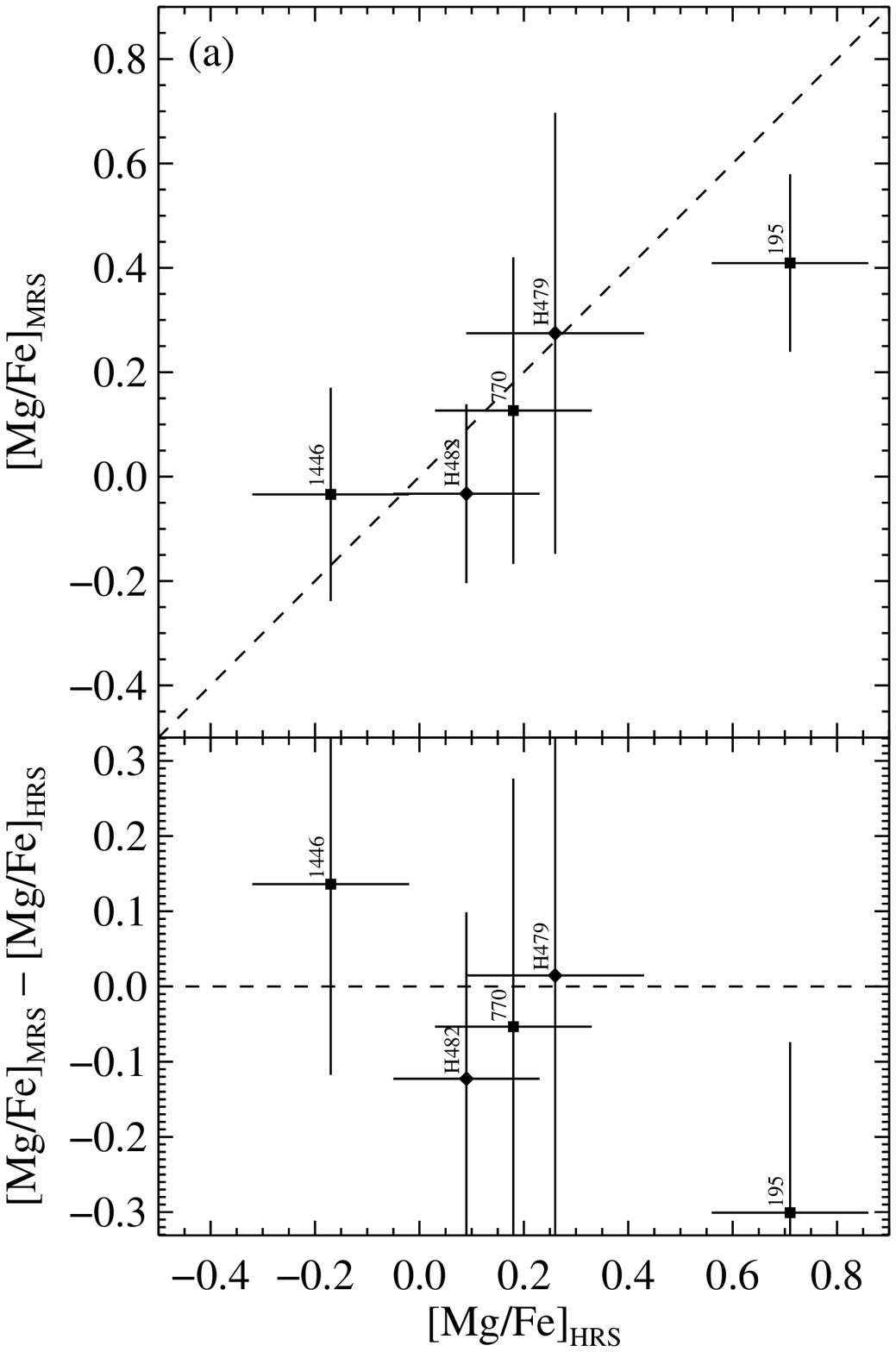}
 \hfil
 \includegraphics[width=0.45\textwidth]{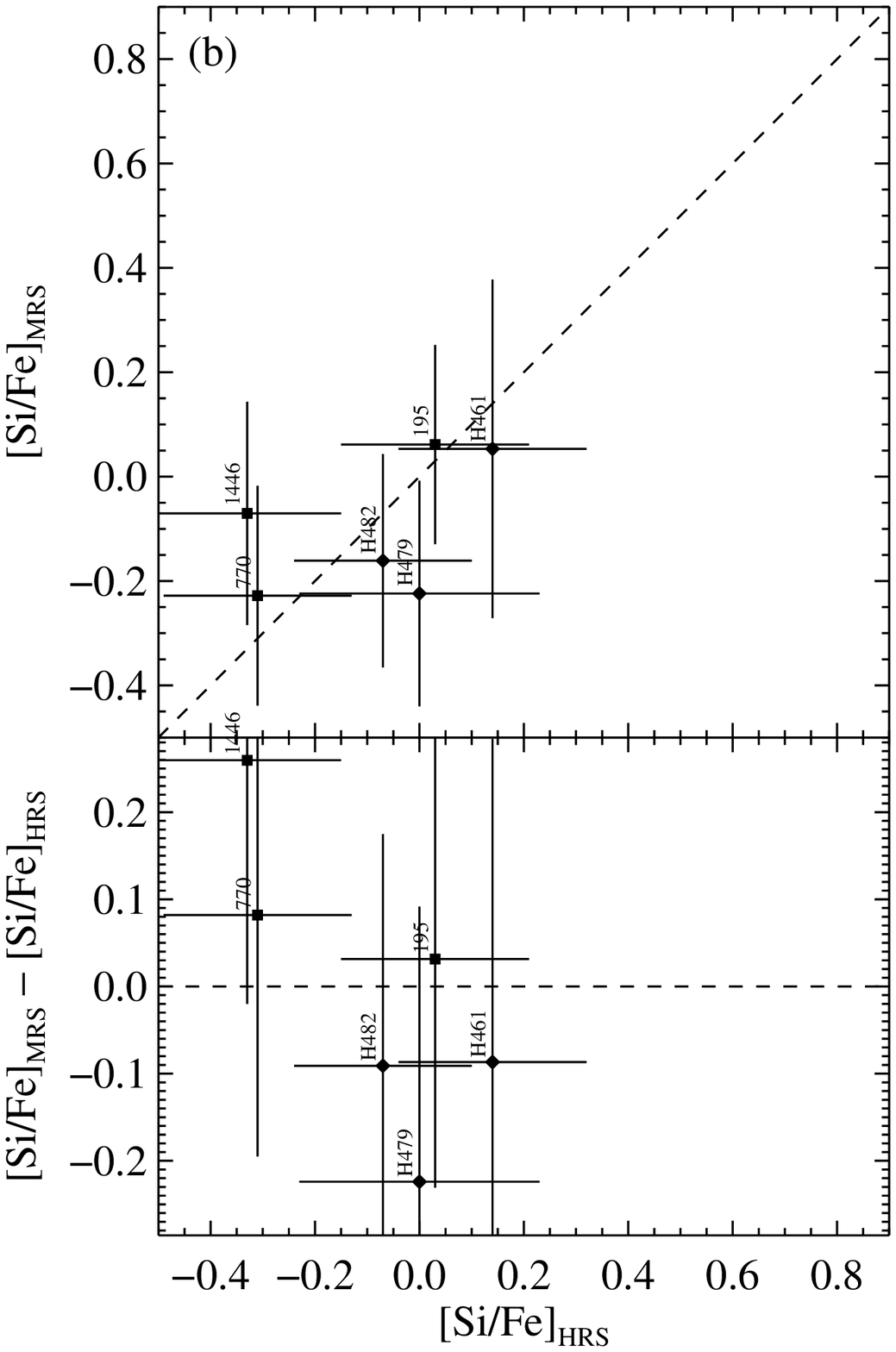}
 \hfil
 \includegraphics[width=0.45\textwidth]{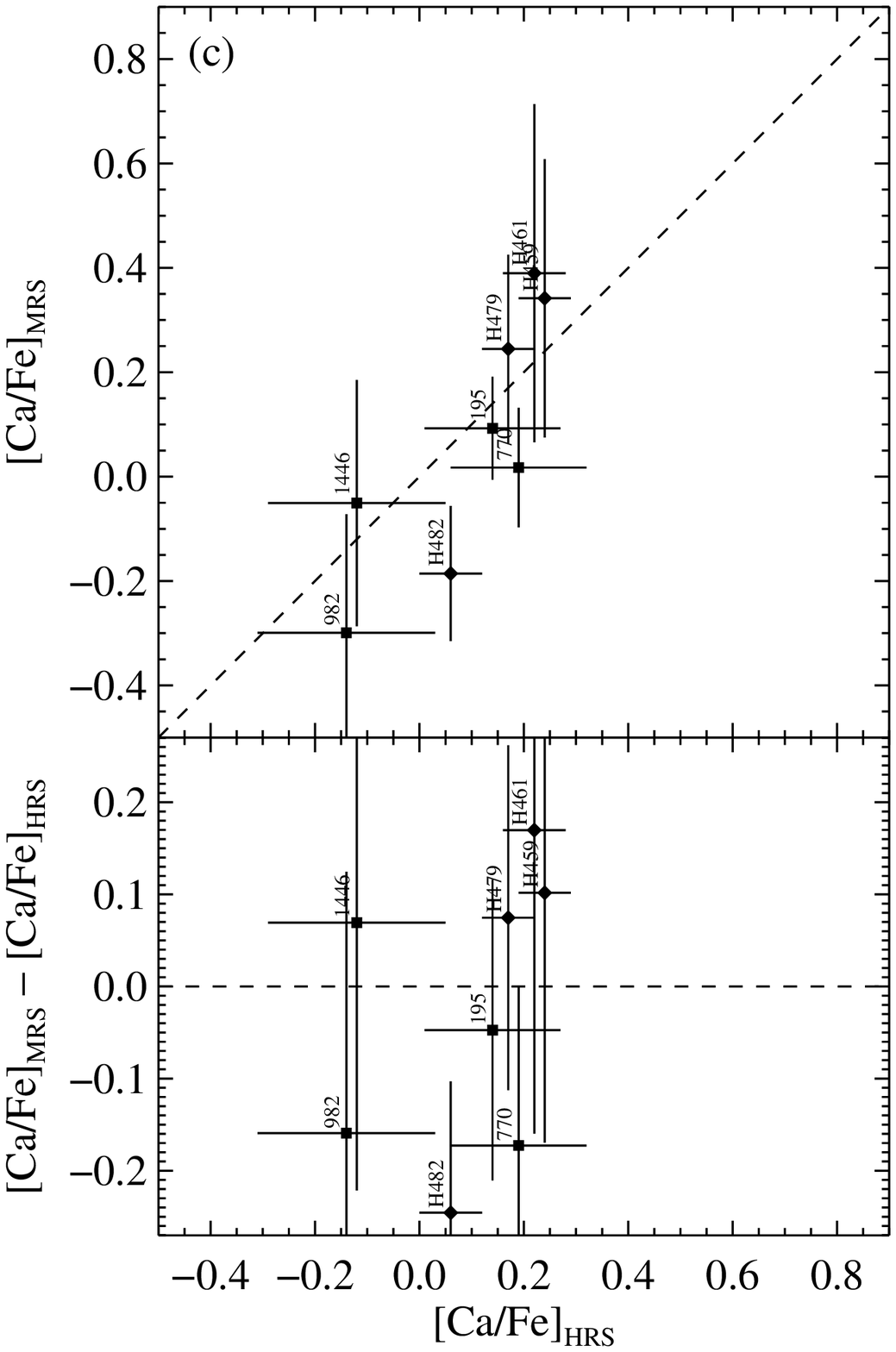}
 \hfil
 \includegraphics[width=0.45\textwidth]{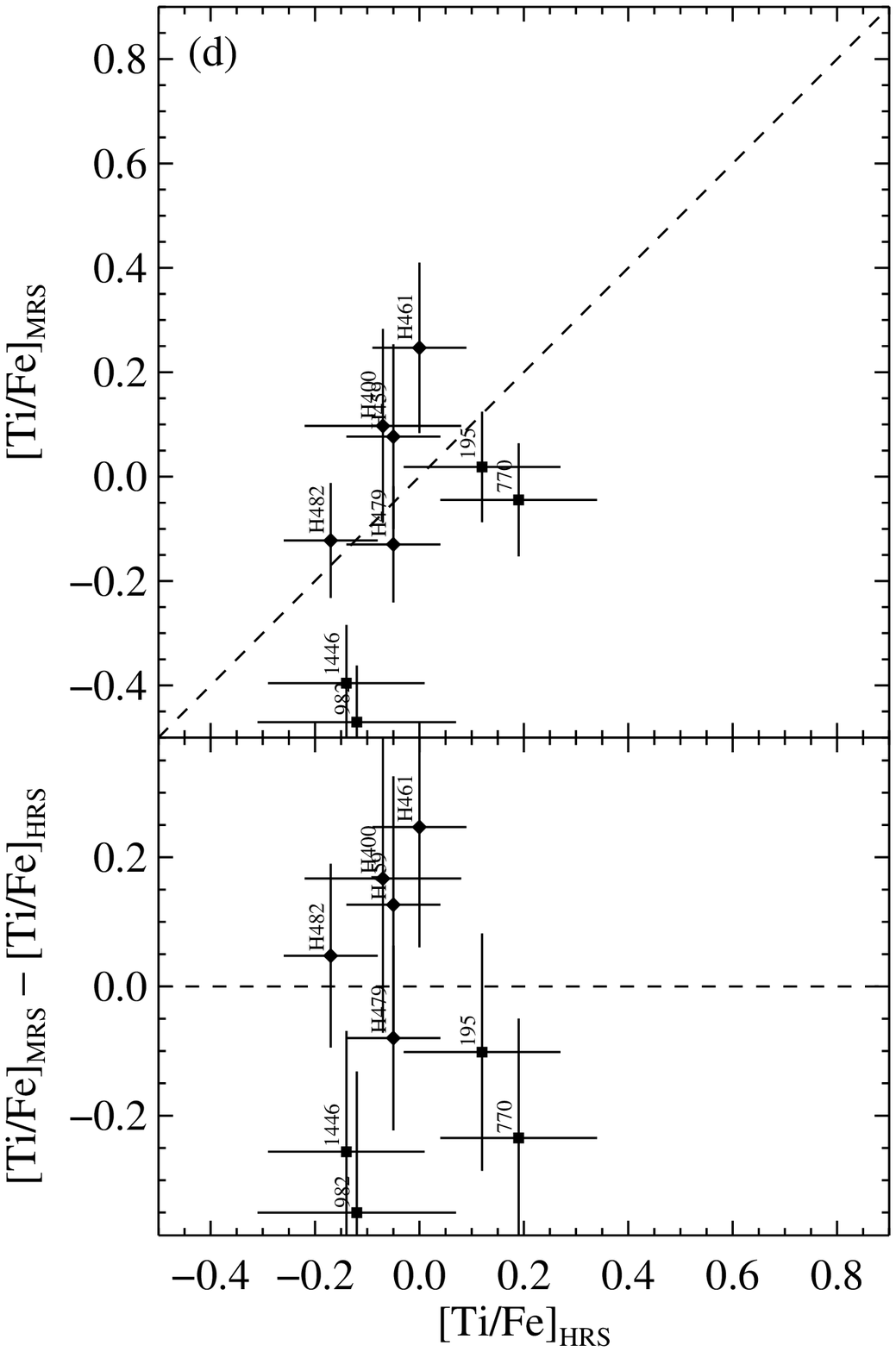}
\caption{Same as Fig.~\ref{fig:hrsfeh} except for [Mg/Fe] ({\it upper
    left}), [Si/Fe] ({\it upper right}), [Ca/Fe] ({\it lower left}),
  and [Ti/Fe] ({\it lower right}).  Only those measurements with
  estimated errors less than 0.45~dex are shown.\label{fig:hrselfe}}
\end{figure}

\begin{figure}[t!]
\centering
\begin{minipage}[t]{0.49\textwidth}
\centering
\includegraphics[width=\linewidth]{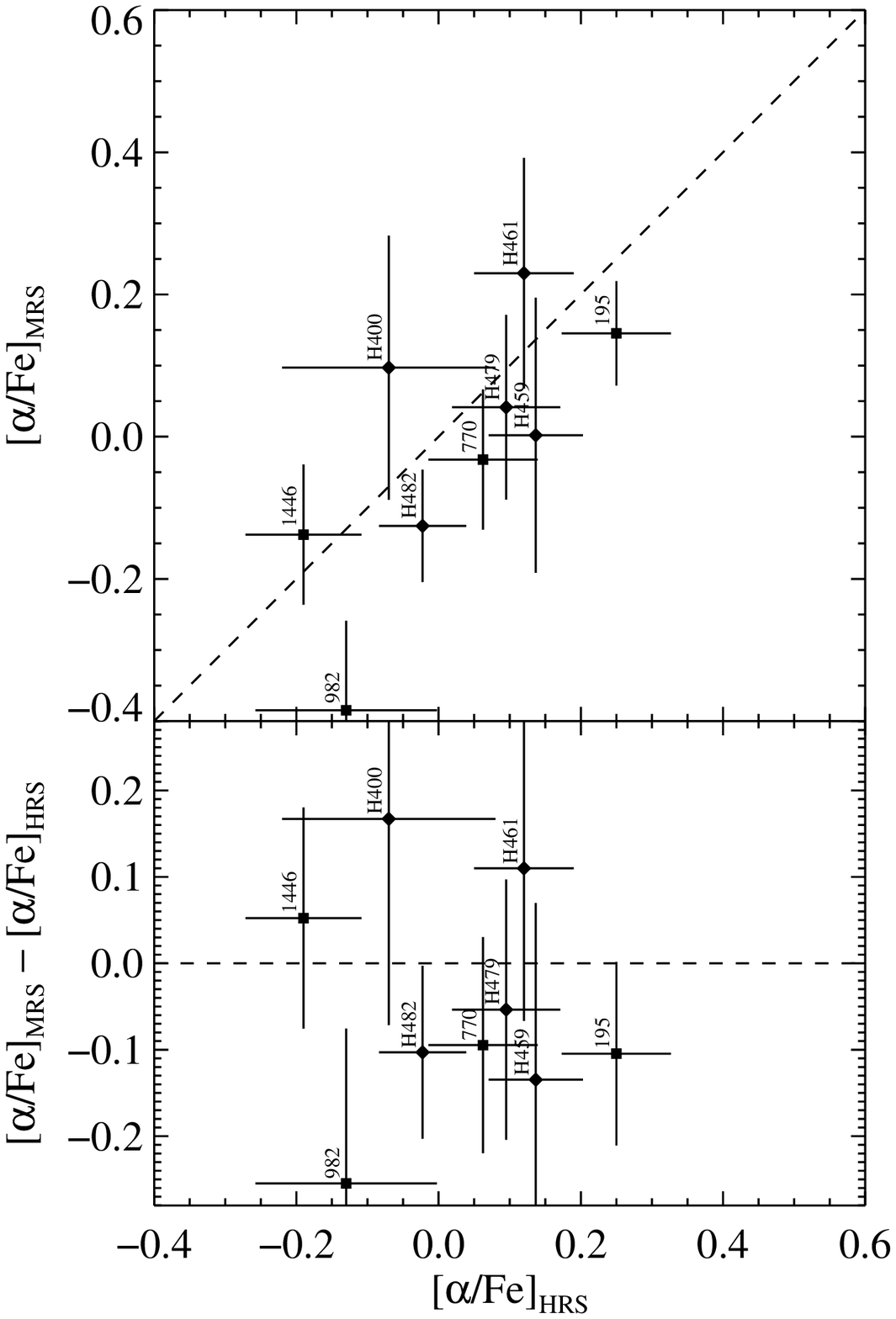}
\caption{Same as Fig.~\ref{fig:hrsfeh} except for an average of the
  $\alpha$ elements.\label{fig:hrsalphafe}}
\end{minipage}
\begin{minipage}[t]{0.49\textwidth}
\centering
\includegraphics[width=\linewidth]{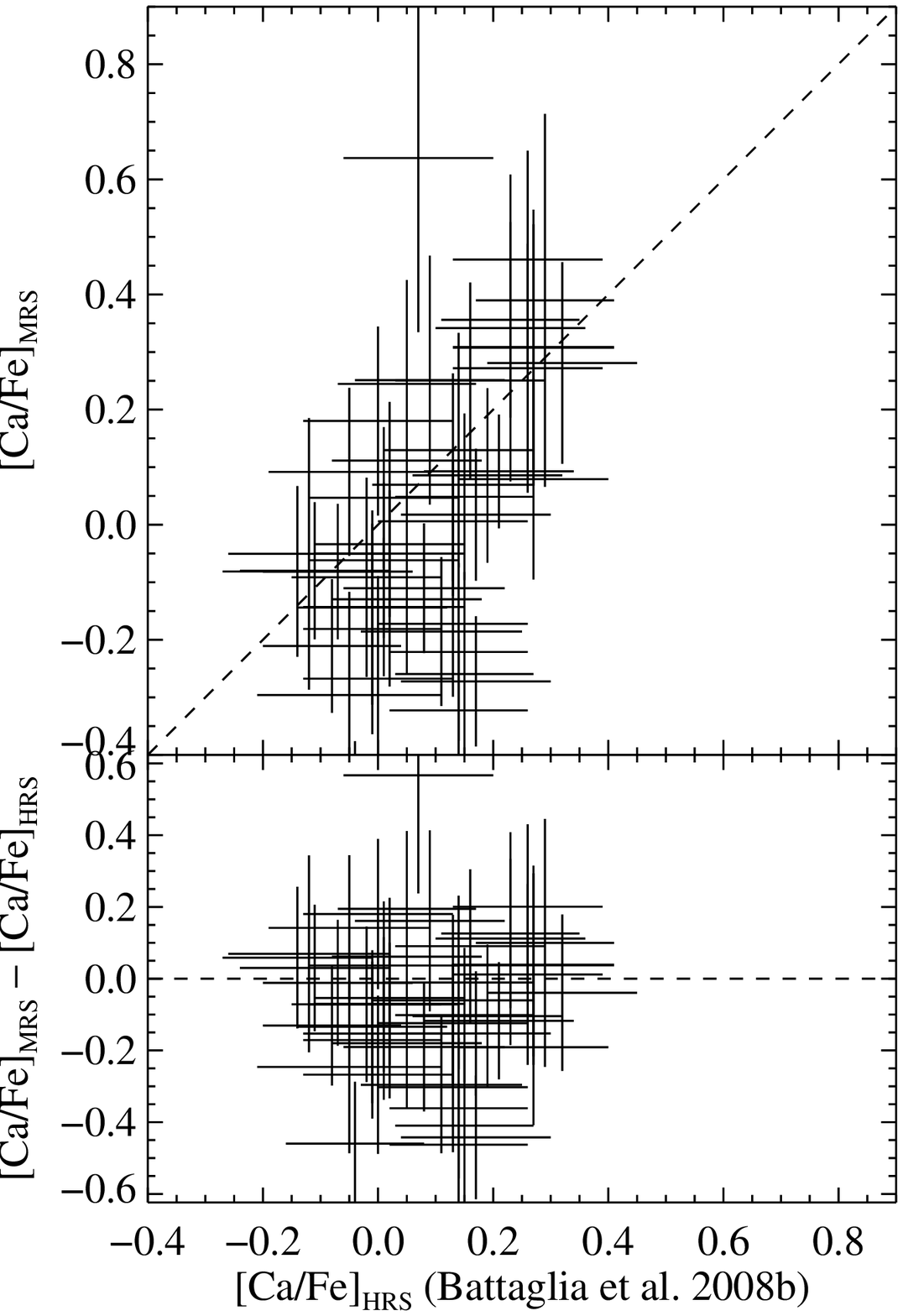}
\caption{Comparison between the HRS measurements of [Ca/Fe] of
  \citet{bat08b} and the synthesis-based medium resolution
  measurements of [Ca/Fe] (this work) for the stars observed in both
  studies.\label{fig:bat08cafe}}
\end{minipage}
\end{figure}

Figure~\ref{fig:hrselfe} shows the comparison between the MRS and HRS
(S03 and G05) values of [Mg/Fe], [Si/Fe], [Ca/Fe], and [Ti/Fe].  In
addition, Fig.~\ref{fig:hrsalphafe} shows unweighted averages of those
four element ratios where available.  The agreement is good in all
cases.  Furthermore, the error bars seem to be reasonable estimates of
the actual random and systematic error.

The agreement between HRS and MRS \afe\ is very good ($\sigma =
\sclalphafehrssigma$~dex).  Even though Fe lines outnumber $\alpha$
elements lines, the ratio \afe\ can be measured about as accurately as
\feh\ because $\alpha$ and Fe respond similarly to errors in
atmospheric parameters whereas \teff\ and \feh\ exhibit strong
covariance.

In addition to \feh, B08b have published HRS measurements of [Ca/Fe].
Figure~\ref{fig:bat08cafe} shows the comparison between the stars we
share in common ($\sigma = \sclcafebatsigma$~dex).  The larger vertical
scatter than horizontal scatter demonstrates that an MRS analysis is
noisier than an HRS analysis when the number of measurable lines is
small.  Regardless, the degree of correlation is high, with a linear
Pearson correlation coefficient of \sclcafebatcorr, indicating that the
medium-resolution spectra have significant power to constrain [Ca/Fe].

\clearpage
\renewcommand{\thetable}{\arabic{table}}
\setcounter{table}{5}
\begin{landscape}
\begin{deluxetable}{ccccccccccccc}
\tabletypesize{\scriptsize}
\tablecolumns{12}
\tablecaption{Multi-Element Abundances in Sculptor\label{tab:data}}
\tablehead{\colhead{RA} & \colhead{Dec} & \colhead{$M$} & \colhead{$T_2$} & \colhead{\teff} & \colhead{\logg} & \colhead{$\xi$} & \colhead{\feh} & \colhead{[Mg/Fe]} & \colhead{[Si/Fe]} & \colhead{[Ca/Fe]} & \colhead{[Ti/Fe]} \\
\colhead{ } & \colhead{ } & \colhead{ } &\colhead{ }  & \colhead{(K)} & \colhead{(cm~s$^{-2}$)} & \colhead{(km~s$^{-1}$)} & \colhead{(dex)} & \colhead{(dex)} & \colhead{(dex)} & \colhead{(dex)} & \colhead{(dex)}}
\startdata
$00^{\mathrm{h}} 59^{\mathrm{m}} 21 \fs 6$ & $-33 \arcdeg 42 \arcmin 58 \arcsec$ & $19.662 \pm 0.090$ & $18.557 \pm 0.048$ & 5105 & 2.02 & 1.66 & $-2.17 \pm 0.19$ &     \nodata      & $-0.12 \pm 0.75$ & $+0.39 \pm 0.28$ &     \nodata      \\
$00^{\mathrm{h}} 59^{\mathrm{m}} 21 \fs 7$ & $-33 \arcdeg 41 \arcmin 02 \arcsec$ & $18.296 \pm 0.037$ & $16.987 \pm 0.027$ & 4674 & 1.27 & 1.84 & $-1.91 \pm 0.14$ & $+0.05 \pm 0.41$ & $+0.07 \pm 0.21$ & $+0.06 \pm 0.15$ & $+0.14 \pm 0.14$ \\
$00^{\mathrm{h}} 59^{\mathrm{m}} 23 \fs 9$ & $-33 \arcdeg 42 \arcmin 59 \arcsec$ & $19.614 \pm 0.091$ & $18.383 \pm 0.048$ & 4805 & 1.87 & 1.70 & $-1.95 \pm 0.16$ & $+0.27 \pm 0.64$ & $-0.12 \pm 0.39$ & $+0.37 \pm 0.22$ & $-0.04 \pm 0.37$ \\
$00^{\mathrm{h}} 59^{\mathrm{m}} 26 \fs 9$ & $-33 \arcdeg 40 \arcmin 29 \arcsec$ & $18.871 \pm 0.051$ & $17.559 \pm 0.029$ & 4672 & 1.49 & 1.79 & $-1.21 \pm 0.14$ & $-0.48 \pm 0.47$ & $-0.06 \pm 0.20$ & $-0.03 \pm 0.14$ & $-0.00 \pm 0.13$ \\
$00^{\mathrm{h}} 59^{\mathrm{m}} 27 \fs 1$ & $-33 \arcdeg 43 \arcmin 42 \arcsec$ & $17.872 \pm 0.026$ & $16.327 \pm 0.024$ & 4392 & 0.88 & 1.93 & $-1.67 \pm 0.14$ & $+0.05 \pm 0.22$ & $-0.19 \pm 0.20$ & $-0.01 \pm 0.13$ & $-0.08 \pm 0.11$ \\
$00^{\mathrm{h}} 59^{\mathrm{m}} 27 \fs 3$ & $-33 \arcdeg 38 \arcmin 47 \arcsec$ & $18.218 \pm 0.037$ & $16.927 \pm 0.028$ & 4738 & 1.26 & 1.84 & $-1.51 \pm 0.14$ & $+0.41 \pm 0.38$ & $-0.32 \pm 0.23$ & $+0.02 \pm 0.15$ & $+0.02 \pm 0.13$ \\
$00^{\mathrm{h}} 59^{\mathrm{m}} 27 \fs 7$ & $-33 \arcdeg 40 \arcmin 35 \arcsec$ & $17.383 \pm 0.025$ & $15.699 \pm 0.025$ & 4232 & 0.60 & 2.00 & $-2.16 \pm 0.14$ & $+0.25 \pm 0.21$ & $+0.14 \pm 0.19$ & $+0.12 \pm 0.12$ & $-0.03 \pm 0.11$ \\
$00^{\mathrm{h}} 59^{\mathrm{m}} 28 \fs 3$ & $-33 \arcdeg 42 \arcmin 07 \arcsec$ & $17.305 \pm 0.017$ & $15.377 \pm 0.020$ & 3789 & 0.49 & 2.03 & $-1.65 \pm 0.14$ & $+0.23 \pm 0.15$ & $+0.30 \pm 0.19$ & $-0.17 \pm 0.13$ & $-0.33 \pm 0.10$ \\
$00^{\mathrm{h}} 59^{\mathrm{m}} 28 \fs 7$ & $-33 \arcdeg 38 \arcmin 57 \arcsec$ & $19.063 \pm 0.037$ & $17.763 \pm 0.030$ & 4622 & 1.58 & 1.77 & $-1.76 \pm 0.14$ &     \nodata      & $+0.07 \pm 0.24$ & $-0.04 \pm 0.16$ & $+0.08 \pm 0.15$ \\
$00^{\mathrm{h}} 59^{\mathrm{m}} 30 \fs 4$ & $-33 \arcdeg 36 \arcmin 05 \arcsec$ & $18.108 \pm 0.024$ & $16.705 \pm 0.021$ & 4517 & 1.11 & 1.88 & $-2.49 \pm 0.15$ &     \nodata      & $-0.19 \pm 1.22$ & $+0.31 \pm 0.21$ & $+0.19 \pm 0.22$ \\
\enddata
\tablecomments{Table~\ref{tab:data} is published in its entirety in
  the electronic edition of the Astrophysical Journal.  A portion is
  shown here for guidance regarding form and content.}
\end{deluxetable}
\clearpage
\end{landscape}

\end{document}